\documentclass[a4paper,fleqn,usenatbib]{mnras}

\usepackage{newtxtext,newtxmath}
 
\usepackage[T1]{fontenc}
\usepackage{ae,aecompl}
\usepackage{graphicx}	
\usepackage{amsmath}	
\usepackage{amssymb}	
\usepackage{diagbox}
\newcommand{\highlight}[1]{{#1}}

\usepackage{soul}

\title[Stronger bars facilitate quenching]{Galaxy Zoo: Stronger bars facilitate quenching in star forming galaxies}

\author[Tobias G\'eron]{
Tobias G\'eron$^{1}$\thanks{E-mail: tobias.geron@physics.ox.ac.uk (TG)},
R. J. Smethurst$^{1}$,
Chris Lintott$^{1}$,
Sandor Kruk$^{2}$,
\newauthor
Karen L. Masters$^{3}$,
Brooke Simmons$^{4}$,
David V. Stark$^{3}$
\\
$^{1}$Oxford Astrophysics, Department of Physics, University of Oxford, Denys Wilkinson Building, Keble Road, Oxford, OX1 3RH, UK\\
$^{2}$ European Space Agency, ESTEC, Keplerlaan 1, NL-2201 AZ, Noordwijk, The Netherlands\\
$^{3}$ Haverford College, Department of Physics and Astronomy, 370 Lancaster Avenue, Haverford, Pennsylvania 19041, USA\\
$^{4}$ Department of Physics, Lancaster University, Lancaster, LA1 4YB, UK\\
}

\date{Accepted XXX. Received YYY; in original form ZZZ}

\pubyear{2021}

\begin{document}
\label{firstpage}
\pagerange{\pageref{firstpage}--\pageref{lastpage}}
\maketitle

\begin{abstract}
    We have used Galaxy Zoo DECaLS (GZD) to study strong and weak bars in disk galaxies. Out of the \highlight{314,000} galaxies in GZD, we created a volume-limited sample (0.01 < z < 0.05, $M_{\textrm{r}} < \highlight{-18.96}$) which contains \highlight{1,867} galaxies with reliable volunteer bar classifications in the ALFALFA footprint. In keeping with previous Galaxy Zoo surveys (such as GZ2), the morphological classifications from GZD agree well with previous morphological surveys. GZD considers galaxies to either have a strong bar (\highlight{15.5\%}), a weak bar (\highlight{28.1\%}) or no bar (\highlight{56.4\%}), based on volunteer classifications on images obtained from the DECaLS survey. This places GZD in a unique position to assess differences between strong and weak bars. We find that the strong bar fraction is typically higher in quiescent galaxies than in star forming galaxies, while the weak bar fraction is similar. Moreover, we have found that strong bars facilitate the quenching process in star forming galaxies, finding higher fibre SFRs, lower gas masses and shorter depletion timescales in these galaxies compared to unbarred galaxies. However, we also found that any differences between strong and weak bars disappear when controlling for bar length. Based on this, we conclude that weak and strong bars are not fundamentally different phenomena. Instead, we propose that there is a continuum of bar types, which varies from `weakest' to `strongest'.
\end{abstract}

\begin{keywords}
galaxies: general -- galaxies: bar -- galaxies: evolution -- galaxies: structure -- galaxies: star formation
\end{keywords}

\section{Introduction}

Since the development of the morphological classification scheme for galaxies \citep{hubble_1926}, spiral galaxies have been divided into two subclasses depending on whether the galaxy was observed to have a bar (SB) or to be unbarred (SA). This scheme is still used \citep{buta_2007, buta_2011}, albeit in modified form \citep{hubble_1936, sandage_1961, devaucouleurs_1959, devaucouleurs_1963, devaucouleurs_1991}. One of the most important changes was the introduction of three subclasses: unbarred (SA), strongly barred (SB) and weakly barred (SAB, \citealt{devaucouleurs_1959, devaucouleurs_1963}). The latter subclass was supposed to be an intermediate class between unbarred and strongly barred, with SAB bars having a length and contrast intermediate between SA and SB bars. Strong bars were typically long and obvious, whereas weak bars are small and faint \citep{devaucouleurs_1959, devaucouleurs_1963}.

It is known that a bar can funnel gas to the centre of galaxy and move angular momentum outwards \citep{sorensen_1976,athanassoula_1992,davoust_2004, athanassoula_2013,villa-vargas_2010,vera_2016,spinoso_2017,george_2019} and thus influence its host galaxy through slow or ``secular'' evolution \citep{kormendy_2004,cheung_2013}. It has been suggested before that a bar could play an important role in quenching its host \citep{kormendy_2004,masters_2011,kruk_2018,fraser_mckelvie_2020b}, although the final mechanism by which the gas is depleted and the galaxy quenches is poorly understood. Studies have shown that the likelihood of hosting a bar increases in more massive, redder and gas-poor galaxies (i.e., typical quiescent galaxies, \citealp{masters_2012,cervantessodi_2017,vera_2016,fraser_mckelvie_2020b}). However, it is unclear whether the bar helps to quench those galaxies or if it is easier to form a bar in a quenched galaxy \citep{masters_2012}.

There are many theories of how bar quenching might proceed. Some suggest that after the inflow of gas to the centre due to the bar, the gas will trigger a starburst. This increases the rate of gas consumption, facilitating the quenching process \citep{alonso_herrero_2001, sheth_2005, jogee_2005, hunt_2008, carles_2016}. Alternatively, the gas can become too dynamically hot for star formation (by increased velocity dispersion or shear), which quenches the host galaxy \citep{zurita_2004, haywood_2016,khoperskov_2018,athanassoula_1992,reynaud_1998,sheth_2000}. 

The simulations of \citet{athanassoula_2013} have demonstrated that bars tend to form later if the gas fraction is high. Some observations show higher SFE/SFR in barred regions \citep{alonso_herrero_2001, hunt_2008, coelho_2011, hirota_2014,janowiecki_2020,magana_serrano_2020, lin_2020}, while others observe lower SFR/SFE \citep{momose_2010,yajima_2019,maeda_2020a}. \citet{sheth_2000} has shown a lower SFR in the region between the centre and bar ends, but higher in the very centre. The observations of \citet{james_2018} suggest a `star formation desert' within the galaxy, where a strong bar suppresses SF by sweeping over it. This is also observed by \citet{spinoso_2017,george_2019,newnham_2020}. These observations demonstrate that bar quenching is localised in the disk region within the bar radius. \citet{watanabe_2011} found similar SFE in the bar and disk region, but an elevated SFE in the bar-end region. These simulations and observations show that barred structures have a definite effect on their host, especially in terms of SFR. 

A reliable and fast method to identify bars and quantify their strength is essential to study them. However, constructing a precise definition of bar strength is not straightforward, even though the idea is intuitive \citep{athanassoula_2003}. Many methods to measure bar strength have been proposed. For example, \citet{devaucouleurs_1959} initially used morphological arguments to differentiate between weak and strong bars; bars in weakly barred galaxies are less evident and shorter, whereas bars in strongly barred galaxies are long and obvious. In the catalogue of detailed visual morphological classifications of \citet{nair_2010}, a bar is defined to be strong if it appears to dominate the light distribution, whereas only a small proportion of the total flux of the galaxy should be found in a weak bar. Other popular methods to measure bar strength include using the m = 2 Fourier mode \citep{athanassoula_2003,garcia_gomez_2017} or using isophotes. It has been shown that both the maximum ellipticity of the bar isophote \citep{athanassoula_1992b,laurikainen_2002,erwin_2004} and the boxiness of the isophotes \citep{gadotti_2011} approximate bar strength. Estimations of the bar torque have also been used before \citep{combes_1981, buta_2001, laurikainen_2002, speltincx_2008}. \citet{hoyle_2011, guo_2019} have found a positive correlation between the bar length and bar strength. Finally, previous work has suggested that weak and strong bars have different surface brightness profiles. Weaker bars have exponential profiles, whereas stronger bars have flatter profiles \citep{elmegreen_1985, elmegreen_1996, kim_2015, kruk_2018}. There are numerous methods to detect and characterize bars, but there is no general consensus among the community on which method is preferable or on the absolute definition of strong and weak bars.

\begin{figure}
	\includegraphics[width=\columnwidth]{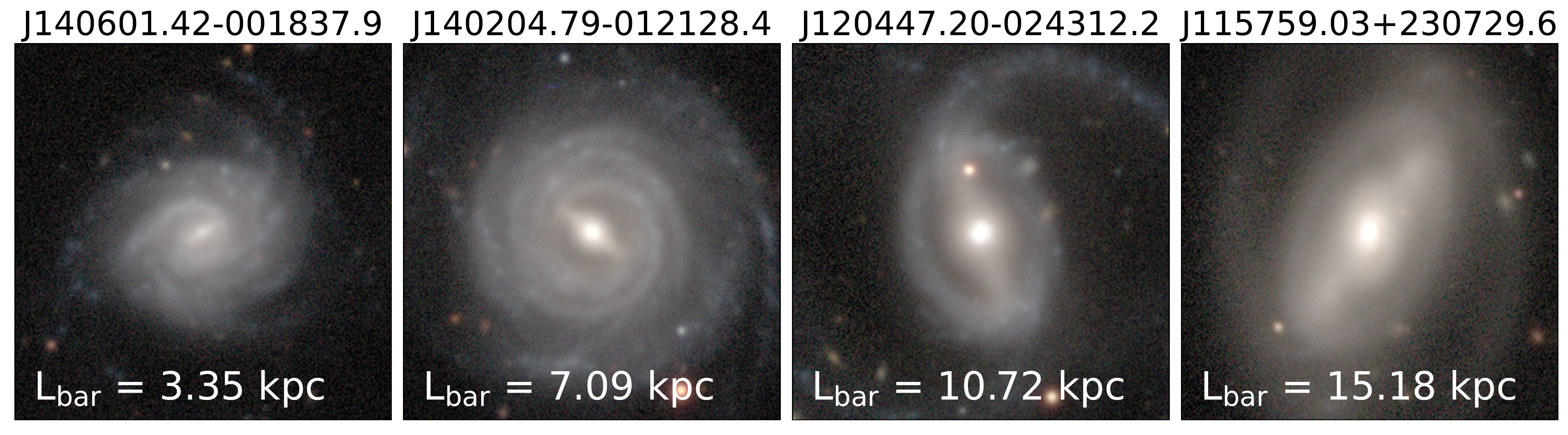}
    \caption{
        DECaLS postage stamps (72x72 arcsec) of low redshift (0.01 < $z$ < 0.025) of galaxies with various bar sizes. See Section \ref{subsection:barlen_measurements} for information about the bar length measurements. This image showcases the variety of bar shapes and sizes.}
    \label{fig:intro_barlen}
\end{figure}

A novel method to approach this problem was investigated in Galaxy Zoo 2 (hereafter GZ2), where citizen scientists were asked to classify SDSS images \citep{york_2000, sdssIV_2017} according to a decision tree structure (please refer to \citet{lintott_2008, lintott_2011, willett_2013} for more details). Whether or not the galaxy had a bar was one question in this decision tree. A total of $\sim$300,000 galaxies were classified using more than 16 million morphological classifications. It has been shown that GZ2 agrees well with morphological catalogues based on expert classifications \citep{willett_2013} and the GZ2 bar classifications have been used to study bars before \citep{hoyle_2011, masters_2011, masters_2012, skibba_2012, cheung_2013, melvin_2014,simmons_2014, cheung_2015, galloway_2015,kruk_2017,kruk_2018,kruk_2019}. Nevertheless, it must be noted that GZ2 only asked whether or not the galaxy had a bar and did not attempt to measure bar strength directly. However, as visualised in Figure \ref{fig:intro_barlen}, bars come in various shapes and sizes, all with varying impacts on their host galaxy. This issue is addressed in Galaxy Zoo DECaLS (GZD), the new version of Galaxy Zoo. In GZD, volunteers chose whether the galaxy had no bar, a weak bar or a strong bar (see Section \ref{subsection:gz}). Images from the Dark Energy Camera Legacy Survey (DECaLS) were used to complement this more detailed question, as DECaLS is a much deeper survey than SDSS (DECaLS has a median $5\sigma$ point source depth of r = 23.6, whereas SDSS DR7 is 95\% complete to r = 22.2, \citealp{abazajian_2009, dey_2019}). DECaLS uses the Dark Energy Camera (DECam, \citealt{flaugher_2015}) on the 4m Blanco telescope at the Cerro Tololo Inter-American Observatory. For a better comparison of the improved imaging and the implications for visual classification, please refer to \citet{walmsley_2021}.

In this paper, we have a statistically robust sample size of both weak and strong bars, identified using Galaxy Zoo DECaLS (GZD). This sample was used to study the differences between weak and strong bars in the context of galaxy evolution, in particular the star formation rate, HI gas mass and depletion timescale. In addition, we aim to answer a more fundamental question: are weak and strong bars intrinsically different physical phenomena?

The structure of the paper is as follows. In Section \ref{section:data} we describe Galaxy Zoo DECaLS in more detail, our data sources and the sample selection. The results are presented in Section \ref{section:results} and discussed in Section \ref{section:discussion}. Finally, we end with our conclusions in Section \ref{section:conclusion}. We assume a standard flat cosmological model using $\rm H_{\rm 0} = 70\;km\,s^{-1}\, Mpc^{-1}$, $\rm \Omega_{\rm m} = 0.3$ and $\rm \Omega_{\rm \Lambda} = 0.7$ where necessary. Photometric quantities were taken from DECaLS \citep{dey_2019}.

\section{Data and sample selection}
\label{section:data}

\subsection{Galaxy Zoo, SDSS and DECaLS}
\label{subsection:gz}

The morphological classifications used in this paper were obtained from the Galaxy Zoo project, where citizen scientists classify galaxies according to a decision tree structure \citep{lintott_2008, lintott_2011}. More specifically, we use the follow-up from Galaxy Zoo 2 (GZ2), namely Galaxy Zoo DECaLS (GZD, \citealp{walmsley_2021}). In GZ2, all images came from the Sloan Digital Sky Survey (SDSS, \citealp{gunn_1998,york_2000,sdssIV_2017}) DR7 \citep{abazajian_2009}, whereas GZD sourced their images from DECaLS \citep{dey_2019}. As DECaLS is a deeper survey (DECaLS has a median $5\sigma$ point source depth of r = 23.6, whereas SDSS DR7 is 95\% complete to r = 22.2, \citealp{abazajian_2009, dey_2019}), this results in some previously invisible features now being visible. GZD only included galaxies that were also in SDSS DR11, which gives us the advantage that we can use preexisting SDSS data of these galaxies, including the Max Planck Institute for Astrophysics and the Johns Hopkins University (MPA-JHU) value added catalogue (VAC) \citep{kauffmann_2003, brinchmann_2004, tremonti_2004}. Nevertheless, this means that the magnitude limit of GZD equals the galaxy completeness limit of SDSS ($m_{\rm r} = 17.77$, \citealp{strauss_2002}). 

A second difference is that GZ2 only asked whether or not a certain galaxy had a bar, while GZD probes the strength of the bar as well by asking the volunteer whether the galaxy has a strong bar, a weak bar or no bar. A detailed explanation of GZD and its decision tree is provided by \citet{walmsley_2021}.

Before moving on, we want to take some time explaining some common nomenclature unique to Galaxy Zoo that will be used frequently throughout this paper. The number of volunteers that answered a particular question, e.g., the bar question, is denoted as $N_{\rm bar}$. The percentage of people that voted for a certain answer for a particular question, e.g., strong bar for the bar question, is denoted as $p_{\rm strong \; bar}$ and is called the vote fraction. Finally, when considering the fraction of strong bars within a population, we denote it as $f_{\rm strong \; bar}$ and call it the strong bar fraction.

\subsection{ALFALFA and upper limits}
\label{subsection:alfalfa}

This paper uses the Arecibo Legacy Fast ALFA (ALFALFA) catalogue of extragalactic HI sources \citep{giovanelli_2005,haynes_2011,haynes_2018} in order to assess HI gas masses. In ALFALFA, the HI gas mass is calculated using: 

\begin{equation}
    \rm{M}_{\rm{HI}} = 2.356 \times 10^{5} \rm{D}^{2}_{\rm{Mpc}} \rm{S}_{\rm{Jy}\, \rm{km} \, \rm{s^{-1}}}\;,
	\label{eq:alfalfa_mass}
\end{equation}

where S is the the integrated HI line flux density in Jy km s$^{-1}$ and D is the distance to the galaxy in Mpc. It is important to note that not all extragalactic sources are detected, as the chance of detection depends on both the integrated HI line flux and the width of the HI profile \citep{haynes_2011,haynes_2018}. This problem is alleviated by estimating an upper limit for these non-detections. This can be done by rearranging Equation 4 of \citet{haynes_2018} to:

\begin{equation}
    \rm{S}_{\rm{Jy}\, \rm{km} \, \rm{s^{-1}}} = SNR \times RMS \times W_{\rm50} / \sqrt{w_{\rm smo}}\;,
	\label{eq:UL_flux}
\end{equation}

where W$_{50}$ is the linewidth of the target (assumed to equal 200 km/s), w$_{\rm smo}$ is a smoothing width and equals W$_{50}/20$, RMS is the RMS noise and SNR is the signal-to-noise ratio. The SNR is assumed to equal 4.5, as the rate of detections in ALFALFA drops rapidly below SNR = 4.5. The RMS is extracted from the spectra, which have a spectral resolution of 10 km s$^{-1}$ \citep{haynes_2018}, at the position of the non-detections. Finally, the HI gas mass upper limit can be computed by substituting the result of Equation \ref{eq:UL_flux} in Equation \ref{eq:alfalfa_mass}. We have assumed that all galaxies without gas measurements that are located within the ALFALFA footprint are non-detections. The footprint for the ALFALFA catalogue of extragalactic HI sources roughly corresponds to $07^{\rm h}30^{\rm m} < \rm{R.A.} < 16^{\rm h}30^{\rm m}$ and $0^{\circ} < \rm{Dec.} < +36^{\circ}$ for the northern hemisphere and $22^{\rm h} < \rm{R.A.} < 03^{\rm h}$ and $0^{\circ} < \rm{Dec.} < +36^{\circ}$ for the southern hemisphere \citep{haynes_2018}, although there is some fringing around the edges.

\subsection{Sample selection}
\label{subsection:sample_selection}

\begin{figure}
	\includegraphics[width=\columnwidth]{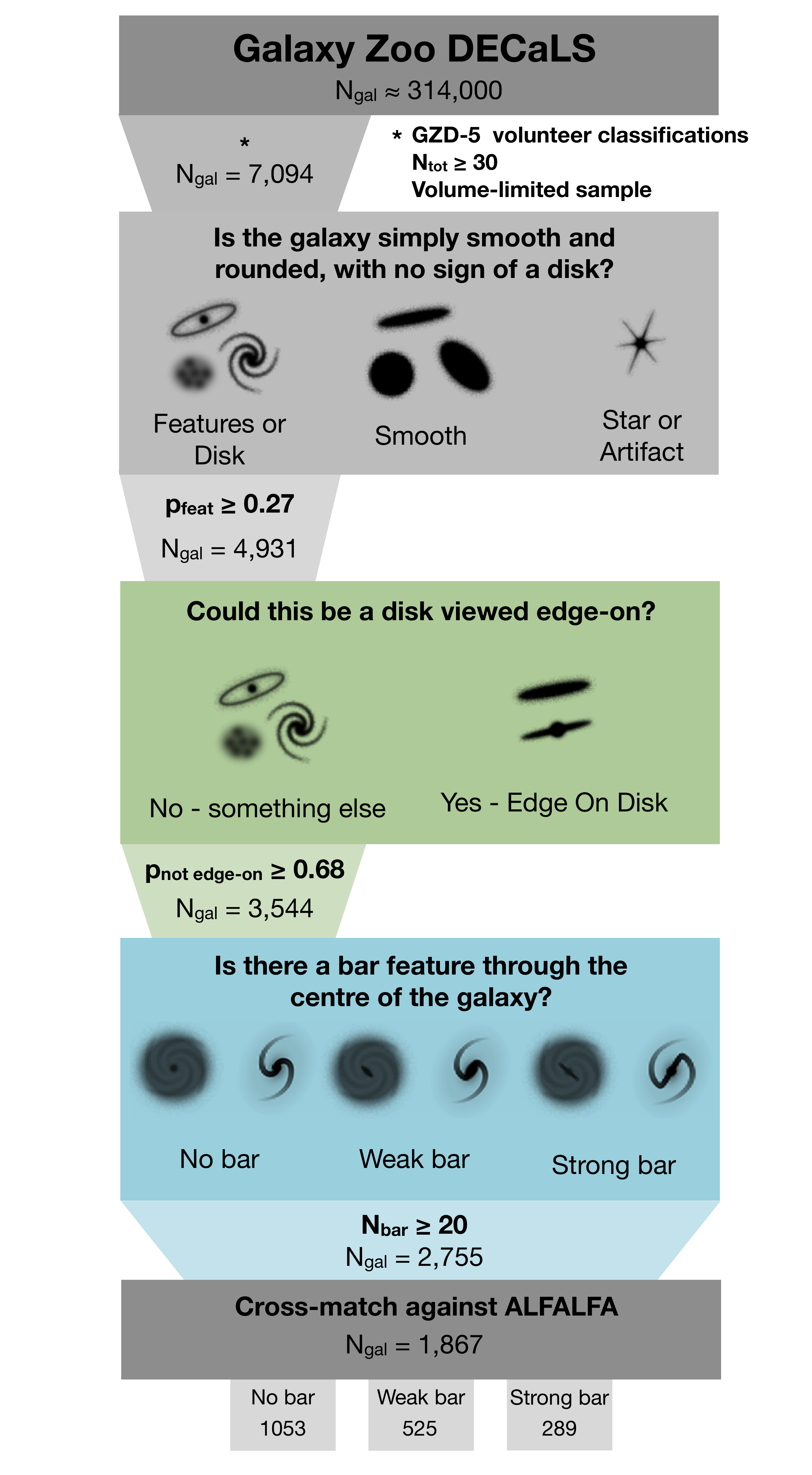}
    \caption{Visualisation of the decision tree structure of GZD, up to the bar question. Note that users will only reach the bar question if they said the galaxy was a disk galaxy that is not viewed edge-on. The thresholds and their effect on the amount of galaxies in our sample are visualised in the funnels between each step. The thresholds on $p_{\textrm{features/disk}}$, $p_{\textrm{not edge-on}}$ and $N_{\textrm{bar}}$ are calculated using the method described in \citet{willett_2013}. Please refer to \citet{walmsley_2021} for the complete decision tree.}
    \label{fig:tree_structure}
\end{figure}

The whole of GZD has $\sim$314,000 galaxies. However, this paper only uses the volunteer classifications from the latest GZD decision tree (GZD-5) and not the machine learning classifications from \citet{walmsley_2021}. We also only considered galaxies with more than \highlight{30} total votes in order to exclusively work with galaxies with the most reliable classifications. This reduced the sample to \highlight{59,337} galaxies. In GZD, galaxies are classified according to a decision tree structure, which means that a volunteer will only answer ``\textit{Is there a bar feature through the centre of the galaxy?}'' after they voted that the galaxy was a disk which is not viewed edge-on. The decision tree is shown (up to the bar question) in Figure \ref{fig:tree_structure}.  

We created a volume-limited sample from these \highlight{59,337} galaxies. The redshift range in our sample is $0.01 < z < 0.05$ in order to match the ALFALFA redshift range. To make this sample volume-limited, the dereddened DECaLS r-band absolute magnitude is constrained to $M_{\textrm{r}} < \highlight{-18.96}$. Additionally, our sample was also cross-matched against the MPA-JHU VAC \citep{kauffmann_2003, brinchmann_2004, tremonti_2004} for stellar mass and (fibre) SFR measurements. This brought the number of galaxies down to \highlight{7,094}. 

As advised in Section 3.3 of \citet{willett_2013}, we also apply limits on the percentage of people that voted that the galaxy has ``features or a disk'' ($p_{\textrm{features/disk}}$), the percentage of people that voted that the galaxy was not edge-on ($p_{\textrm{not edge-on}}$) and the amount of people that voted in the bar question ($N_{\textrm{bar}}$). Thresholds on these parameters are shown to limit our sample to not edge-on disk galaxies that may or may not contain bars. However, as the thresholds quoted in \citet{willett_2013} were calculated using the GZ2 dataset, we have to compute new values based on GZD using the method described in \citet{willett_2013}. These new values are: $p_{\textrm{features/disk}} \geq \highlight{0.27}$ and $p_{\textrm{not edge-on}} \geq \highlight{0.68}$ for $N_{\textrm{bar}} \geq \highlight{20}$, which brought the total number of galaxies with reliable classifications in our sample to \highlight{2,755}. The effects of these cuts are shown in detail in Figure \ref{fig:tree_structure}. Although we only require $p_{\textrm{features/disk}} \geq \highlight{0.27}$, in practice, the value for $p_{\textrm{features/disk}}$ will be much higher. This is because most galaxies have a total of 40 unique classifications and we also limit our sample to galaxies that have $N_{\textrm{bar}} \geq \highlight{20}$. For reference, the median $p_{\textrm{features/disk}}$ of our final sample is equal to \highlight{0.82} and the 10$^{\rm th}$ percentile is still \highlight{0.58.}

It is worth pointing out that the bar classifications might be affected by inclination. Even though we filter out the most edge-on galaxies, it is possible that the same bar in a face-on galaxy will be perceived differently by volunteers in a more inclined galaxy. Thus, a more stringent inclination threshold was considered. However, there is a trade-off between clean samples and sample size. We opted for a bigger sample size and judged that our threshold of $p_{\textrm{not edge-on}} \geq \highlight{0.68}$ is sufficient for our purposes.

The last step in our sample selection is cross-matching against ALFALFA to obtain HI gas mass measurements \citep{giovanelli_2005,haynes_2011,haynes_2018}, which reduced the final sample size to \highlight{1,867} galaxies.

\begin{table}
    \centering
    \caption{Every galaxy is assigned a bar type (strong bar, weak bar and no bar) based on their respective vote fractions ($p_{\textrm{strong bar}}$, $p_{\textrm{weak bar}}$ and $p_{\textrm{no bar}}$) according to the following scheme. \vspace{0.2cm}}
    \label{tab:bar_conditions}
    \begin{tabular}{ccl}
    \hline
    Condition 1 & Condition 2 & Result \\
    \hline
    $p_{\textrm{strong bar + weak bar}} < 0.5$ & N/A & No bar\\
    $p_{\textrm{strong bar + weak bar}} \geq 0.5$ & $p_{\textrm{strong bar}} < p_{\textrm{weak bar}}$ & Weak bar\\
    $p_{\textrm{strong bar + weak bar}} \geq 0.5$ & $p_{\textrm{strong bar}} \geq p_{\textrm{weak bar}}$ & Strong bar\\
    \hline
    \end{tabular}
\end{table}

\begin{figure*}
	\includegraphics[width=\textwidth]{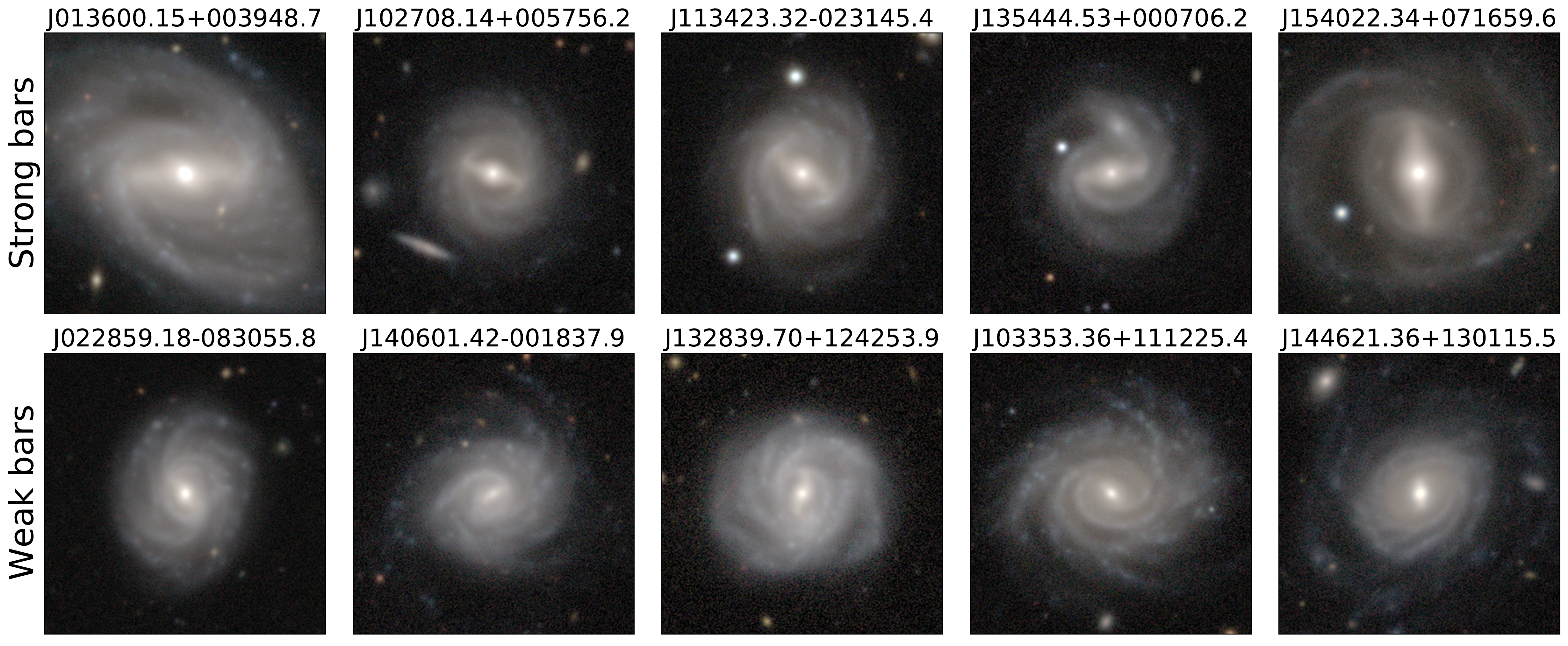}
    \caption{
        DECaLS postage stamps (72x72 arcsec) of five randomly selected galaxies with high Petrosian radii (> 18 arcsec) and high strong bar vote fractions ($p_{\textrm{strong bar}}$ > 0.6, top row) and weak bar vote fractions ($p_{\textrm{weak bar}}$ > 0.6, bottom row) in GZD. Notice the difference in bar length and strength between the strong and weak bars.}
    \label{fig:SBWB}
\end{figure*}

Every galaxy was assigned one of three bar types (strong, weak or no bar) using the following scheme: if $p_{\textrm{strong bar + weak bar}} < 0.5$, the galaxy had no bar. If the previous condition was not met and if $p_{\textrm{strong bar}} \geq p_{\textrm{weak bar}}$, then the galaxy had a strong bar. If not, it had a weak bar. This scheme is summarized in Table \ref{tab:bar_conditions}. This resulted in \highlight{15.5\%} of our galaxies (\highlight{289/1,867}) having a strong bar, \highlight{28.1\%} (\highlight{525/1,867}) having a weak bar and \highlight{56.4\%} (\highlight{1,053/1,867}) having no bar. Thus, the total barred fraction equals \highlight{43.6\%}, in general agreement with the literature \citep{devaucouleurs_1991,eskridge_2000,menendez-delmestre_2007, barazza_2008,aguerri_2009,nair_2010,masters_2011,buta_2019,zhao_2020}, although we study and compare bar fractions in more detail in Sections \ref{subsubsection:comparison_gz2} and \ref{subsection:demographics}. Figure \ref{fig:SBWB} shows random examples of galaxies with a strong or a weak bar.

\subsection{Bar length measurements}
\label{subsection:barlen_measurements}
In an analysis of bar types, bar length measurements are desirable. A citizen science project aimed at measuring bar lengths for GZ2 galaxies was previously carried out by \citet{hoyle_2011}\footnote{The bar length values in \citet{hoyle_2011} are quoted in units of kpc/h. However, closer inspection revealed that this is an error, and the bar lengths given by \citet{hoyle_2011} are actually in kpc. We have made this correction throughout this work.}. However, only $\sim$\highlight{22}\% of the galaxies in our volume-limited sample had bar length measurements in \citet{hoyle_2011}. To alleviate this issue, one of the authors (TG) measured the bar length of all galaxies with a weak or strong bar in our volume-limited sample of GZD (a total of \highlight{814} bars). Of those, \highlight{177} also appeared in \citet{hoyle_2011}. An additional \highlight{60} galaxies from \citet{hoyle_2011} (that were not in GZD) were measured as well to better compare the two bar length catalogues. This brought the total amount of bar length measurements to \highlight{874}. Here, bar length is defined as the length of the entire bar (i.e., twice the bar radius). Every galaxy was measured twice and the average length (in arcsec) was taken. The order of the measurements was randomized, so that it was unknown while measuring whether the bar was classified by GZD as weak or strong. We used the angular diameter distance (which was calculated using SDSS redshifts) to convert these bar lengths from arcsec into kpc.

\begin{figure}
	\includegraphics[width=\columnwidth]{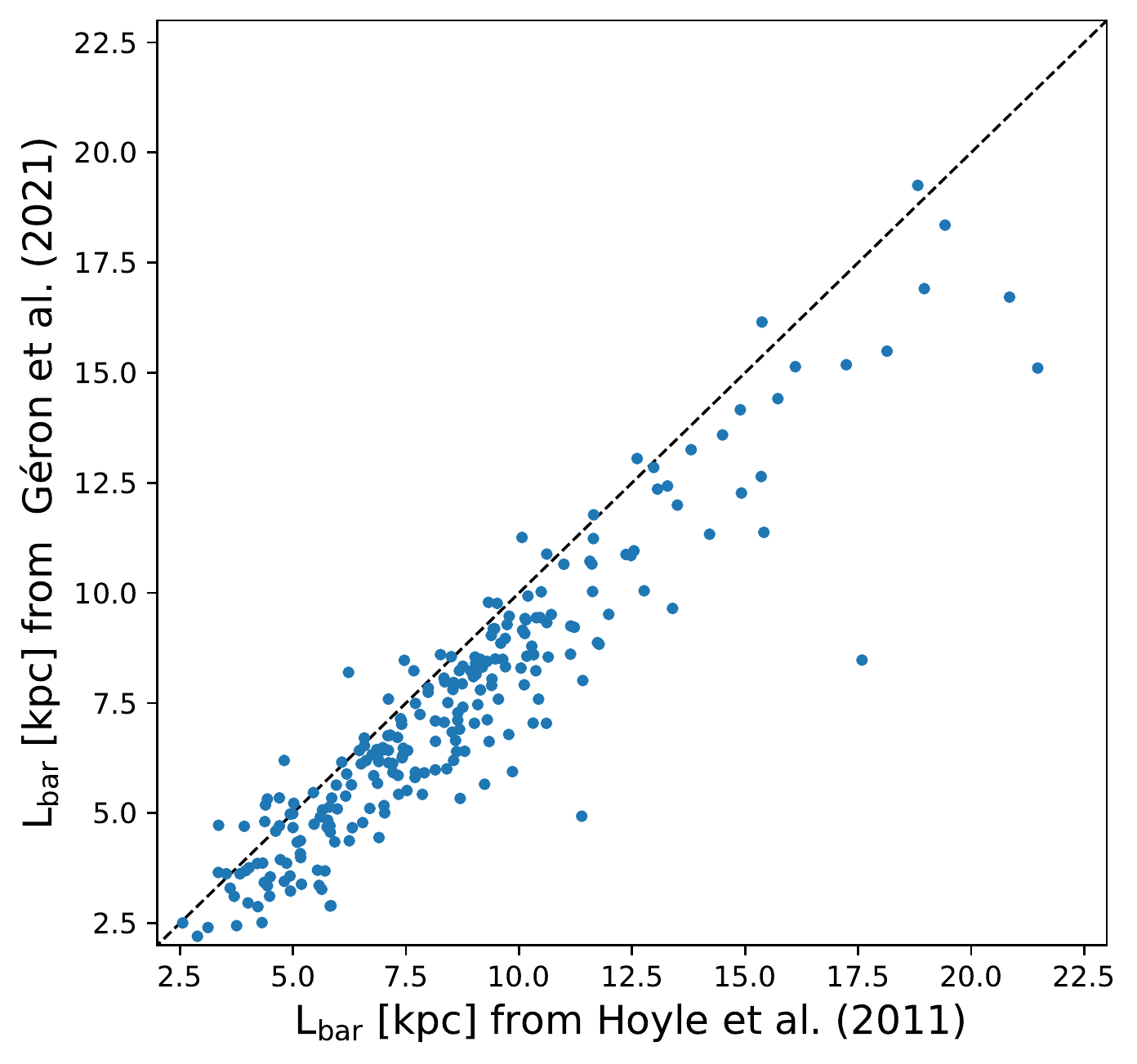}
    \caption{A comparison between the bar lengths measured in this study and those of \citet{hoyle_2011} for \highlight{237} cross-matched galaxies. The dashed line indicates where the measurements would be identical. Our measurements are, on average, \highlight{87.8\%} the size of those of \citet{hoyle_2011}. This difference is the result of different imaging used: \citet{hoyle_2011} used SDSS DR6, while we had access to deeper DECaLS images. See Section \ref{subsection:barlen_measurements} for details on how these bar lengths were measured.}
    \label{fig:comp_barlen}
\end{figure}

Our bar length measurements and those of \citet{hoyle_2011} agree fairly well (Figure \ref{fig:comp_barlen}). On average, our bar length measurements are \highlight{87.8}\% the size of those in \citet{hoyle_2011}. This difference is likely to result from us having access to deeper and more detailed DECaLS images with a smaller PSF, whereas \citet{hoyle_2011} used SDSS DR6 \citep{adelman_mccarthy_2008} images on a Google Maps interface. 

We also compute the relative bar length, obtained by dividing the length of the bar by the diameter of the galaxy. We opted to use the well-established Petrosian radii from SDSS-IV DR16 to calculate the Petrosian diameter \citep{sdssIV_2017, ahumada_2020}.

\subsection{Comparison to other catalogues}
\subsubsection{Galaxy Zoo 2}
\label{subsubsection:comparison_gz2}

It is important to verify whether different versions of Galaxy Zoo are consistent with each other. Therefore, we compare the results of GZD with GZ2. In order to do this, we cross-matched our sample with GZ2, using the appropriate cuts for GZ2 (see \citealt{willett_2013}). A total of \highlight{1,160} galaxies overlap in both samples. Note that, as we are using raw vote fractions in GZD, we are comparing them to the raw vote fractions of GZ2 in order to compare like with like. For more information on GZ2 raw and debiased vote fractions, please refer to \citet{willett_2013} and \citet{hart_2016} and to \citet{walmsley_2021} for a broader comparison of GZ2 with GZD.

\begin{figure*}
	\includegraphics[width=\textwidth]{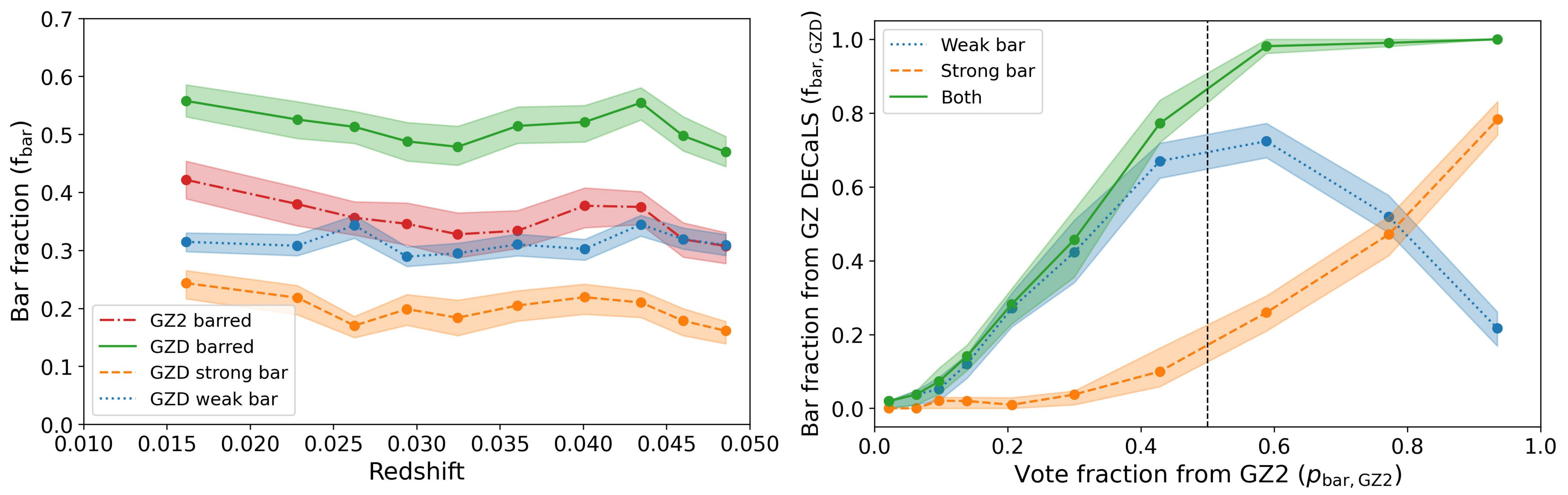}
    \caption{A comparison between Galaxy Zoo 2 \citep{willett_2013} and GZD. \textbf{Left}: The various bar fractions (from GZD and GZ2) over redshift for the galaxies in that are in both catalogues. All the galaxies are binned in equal-sized bins. \textbf{Right}: The various GZD bar fractions (f$_{\textrm{bar, GZD}}$) compared to the GZ2 vote fraction ($p_{\textrm{bar, GZ2}}$). The contours represent the 3$\sigma$ region after bootstrapping the data 10,000 times and retaining 90\% of the data for each iteration. The vertical dashed line represents the threshold many GZ2 studies use to select bars ($p_{\textrm{bar, GZ2}} > 0.5$). It is clear that GZD identifies many more barred structures than GZ2. Most of the `newly barred' galaxies in GZD appear have a weak bar.}
    \label{fig:gzd_gz2_comb}
\end{figure*}

In the left panel of Figure \ref{fig:gzd_gz2_comb}, the GZ2 bar fraction, as well as the GZD weak bar fraction, the GZD strong bar fraction and the GZD total bar fraction are visualised over redshift. GZ2 bars were identified by using a $p_{\textrm{bar, GZ2}}$ threshold of 0.5, similarly to many previous GZ2 studies \citep{masters_2011,masters_2012, cheung_2015, kruk_2017,kruk_2018}. The GZD bars were identified according to the scheme described in Table \ref{tab:bar_conditions}. We can very clearly see that we identify many more bars with GZD. In fact, we find as many weak bars in GZD as there were bars in GZ2. This can be due to either deeper imaging (provided by DECaLS) or due to the different decision trees employed. It is possible that in GZ2, because volunteers were not given an intermediate option, they defaulted to a ‘no bar’ classification when unsure. 

If the latter effect plays a significant role in our increased observed bar fraction, then many of our GZD galaxies with a weak bar will have low to average GZ2 bar vote fractions ($p_{\textrm{bar, GZ2}}$). This is indeed what we observe in the right panel of Figure \ref{fig:gzd_gz2_comb}. Here, the raw GZ2 vote fraction is plotted against the GZD bar fraction. As an illustration, the galaxies with a GZ2 vote fraction of $\sim$0.4 have an average GZD weak barred fraction of $\sim$0.6. As noted before, many previous GZ2 studies used $p_{\textrm{bar, GZ2}} = 0.5$ as a threshold to select bars \citep{masters_2011,masters_2012, cheung_2013, cheung_2015, kruk_2017,kruk_2018}. Here, we see that the strong bar fraction is very low in the $p_{\textrm{bar, GZ2}} < 0.5$ range, suggesting that the vast majority of the GZD galaxies with a strong bar were already picked up by GZ2, while most newly found barred galaxies in GZD have a weak bar. Previous GZ2 studies usually noted that their barred sample mainly consisted of strongly barred galaxies \citep{masters_2011, cheung_2013, cheung_2015, kruk_2017, kruk_2018}. For example, \citet{masters_2012} found that >90\% of the strong and intermediate bar types from \citet{nair_2010} have $p_{\textrm{bar, GZ2}} > 0.5$. This notion is now verified by GZD as well and suggests that results from previous GZ2 studies mainly apply to strongly barred galaxies, not weakly barred galaxies.

The weak bar fraction rises from $\sim$0.0 up to $\sim$0.7 in the $p_{\textrm{bar, GZ2}} < 0.5$ range. In addition, the GZD weak bar fraction is still quite high in the $p_{\textrm{bar, GZ2}} > 0.5$ range (at a GZ2 vote fraction of $\sim$0.8, the GZD weak bar fraction is still $\sim$0.4). These galaxies are thought to have a very clear weak bar, which is not necessarily the same as having a strong bar. Examples of very clear weak bars are shown in the bottom row of Figure \ref{fig:SBWB}. Finally, the GZD strong bar fraction and GZD total bar fraction correlate nicely with the GZ2 vote fraction.

In conclusion, the GZD total barred vote fraction is higher than the GZ2 barred vote fraction due to deeper imaging and an improved decision tree. The GZD strong bar fraction correlates with the GZ2 barred vote fraction. Most GZD strong bars were already identified in GZ2, while weak bars make up the bulk of the newly detected bars in GZD, compared with GZ2.

\subsubsection{Nair and Abraham}
\label{subsubsection:na}

A comparison of our GZD sample with the catalogue of visual morphological classifications of \citet{nair_2010} is shown in Table \ref{tab:gzdna}. A  total of \highlight{589} galaxies appear in both samples. GZD classifies galaxies into having no bar, a weak bar or a strong bar, whereas \citet{nair_2010} has many more options: strong, intermediate, weak, ansae, peanut, nuclear, unsure or no bar. However, the ansae, peanut, nuclear and unsure categories are rare and were combined into one category called `others' in our comparison. It is also worth noting that a single galaxy can have multiple classifications in \citet{nair_2010}, but this did not occur frequently. 

A total of \highlight{25} out of \highlight{27} galaxies classified by \citet{nair_2010} as being strongly barred are also identified in GZD as having a strong bar. Also, \highlight{306} out of \highlight{338} of the galaxies we classified as having no bar were also classified by \citet{nair_2010} as being unbarred. Not a single galaxy we classified as having no bar was classified as strongly barred by \citet{nair_2010}. However, there is a small subset of galaxies (\highlight{23} in total) that volunteers classified as having a strong bar in GZD, while they were identified as being unbarred in \citet{nair_2010}. The authors of this paper visually inspected those galaxies separately and concluded that the most of them do have a strong bar (see Appendix \ref{app:na}). A possibility for this apparent contradiction is that \citet{nair_2010} classified galaxies based on SDSS images, while the volunteers got shown deeper and more detailed DECaLS images.

\begin{table}
    \scriptsize
    \caption{A comparison between GZD and the catalogue of visual morphological classifications of \citet{nair_2010}. \textbf{Top row}: GZD galaxies with a strong bar. \textbf{Middle row}: GZD galaxies with a weak bar. \textbf{Bottom row}: GZD galaxies without bars. Each cell indicates the percent of GZD bars that are classified by \citet{nair_2010} as having a given feature, with the sample size in parentheses. E.g.: \highlight{21.4}\% of the GZD galaxies with a strong bar are also classified as strongly barred by \citet{nair_2010}.}
    \label{tab:gzdna}
    \begin{tabular}{c c c c c c }
     \hline
     \diagbox[width = 1.4cm]{GZD}{NA10}& Strong bar & Intermediate bar & Weak bar & No bar & Others \\
     \hline
     Strong bar & \textbf{21.4\% (25)} & \textbf{43.6\% (51)} & 8.5\% (10) & 19.7\% (23) & 6.8\% (8)\\
     Weak bar & 1.3\% (2) & \textbf{27.1\% (42)} & \textbf{23.9\% (37)} & \textbf{42.6\% (66)} & 5.2\% (8)\\
     No bar & 0.0\% (0) & 1.2\% (4) & 5.3\% (18) & \textbf{90.5\% (306)} & 3.0\% (10)\\
     \hline
    \end{tabular}
\end{table}

\section{Results}
\label{section:results}

\subsection{Demographics of galaxies hosting bars}
\label{subsection:demographics}
\subsubsection{Properties of galaxies hosting weak and strong bars}
\label{subsubsection:prop_gals}

\begin{figure*}
	\includegraphics[width=\textwidth]{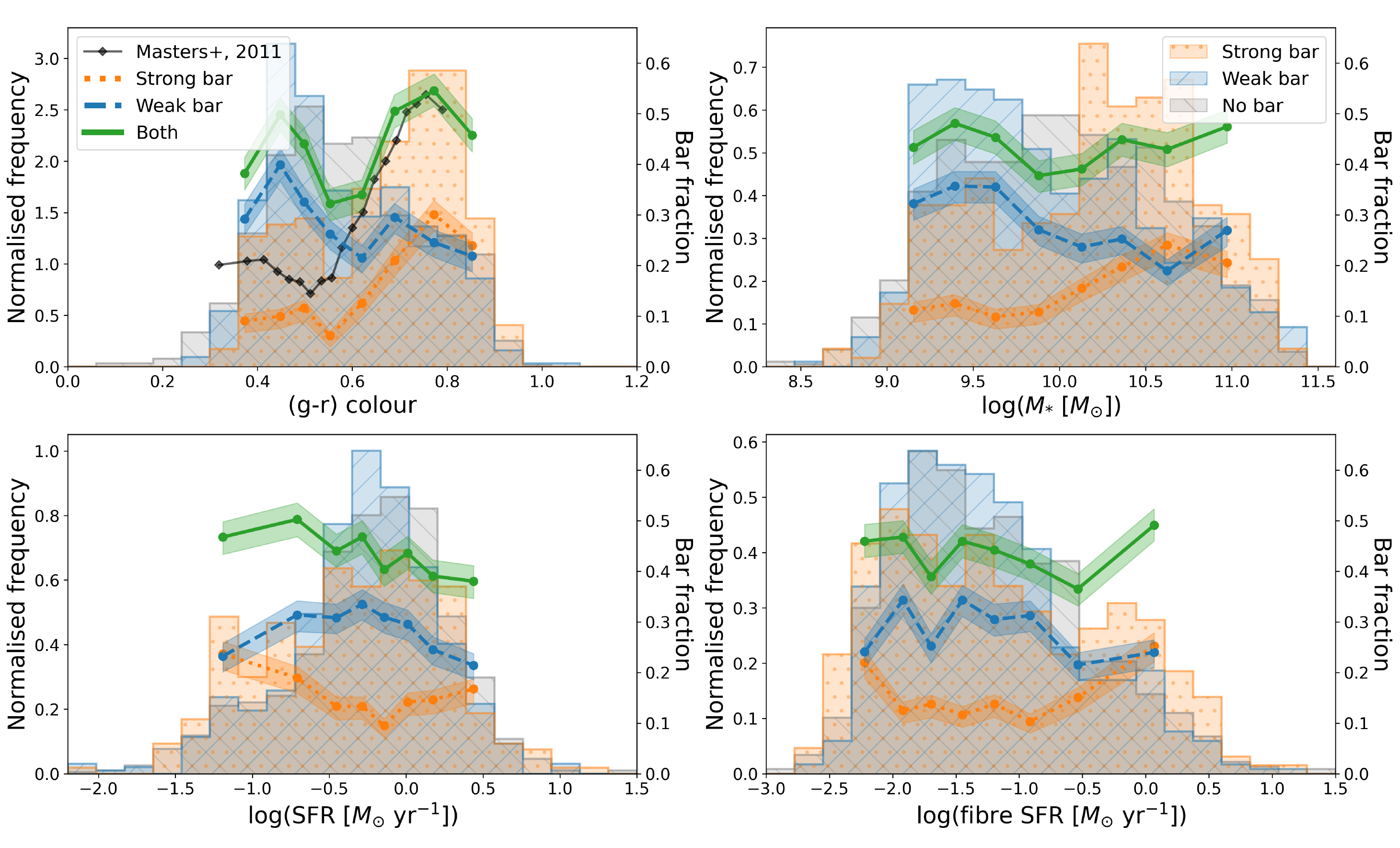}
    \caption{\textbf{Top left}: The effect of (g-r) colour on weak, strong and combined bar fraction (secondary y-axis). All the galaxies are binned in equal-sized bins. The contours represent the 3$\sigma$ region after bootstrapping the data 10,000 times and retaining 90\% of the data for each iteration. The histograms in the background show the normalised frequencies of galaxies with a strong bar, a weak bar and no bar (primary y-axis). \textbf{Top right}: The effect of stellar mass on the various bar fractions. \textbf{Bottom left}: The effect of SFR. \textbf{Bottom right}: The effect of fibre SFR. This figure shows that strong bars drive the consensus that bars appear more often massive, red and quiescent galaxies. The weak bar fraction actually decreases with colour and stellar mass. Magnitudes obtained from SDSS and SFRs, fibre SFRs and stellar masses obtained from MPA-JHU.}
    \label{fig:barfrac_plots}
\end{figure*}

Previous studies have found that bars are more likely to be found in massive, red, quiescent galaxies \citep{masters_2012,cervantessodi_2017}, but they did not distinguish between weak and strong bars. In Figure \ref{fig:barfrac_plots}, we show how the GZD galaxy population with a weak or strong bar changes in terms of (g-r) colour, stellar mass, global SFR and fibre SFR. The latter three parameters were taken from the MPA-JHU catalogue \citep{kauffmann_2003,brinchmann_2004,tremonti_2004}. The reason we included fibre SFR in this analysis is because this probes the SFR in the central 3 arcsecs of the galaxy. As bars are thought to funnel gas to the centre and increase SFR there, this region is of most interest to determine the effect of the bar. However, the physical size that the 3 arcsec region of the fibre probes will change depending on the redshift of the galaxy. For our sample, it varies between \highlight{0.69} and \highlight{2.93} kpc. However, as we expect no changes in the proportion of bar types over the small redshift range we are probing, this effect is negligible as it will affect the three groups equally. For more details, please refer to Appendix \ref{app:fibre}.

The distribution of the bar types for every parameter in Figure \ref{fig:barfrac_plots} is shown in terms of bar fractions (the lines, secondary y-axis) and counts (histograms, normalised so that the area under each histogram equals one, primary y-axis). The top-left plot of Figure \ref{fig:barfrac_plots} shows the (g-r) colour. We see that the strong bar fraction increases at redder colours, in agreement with the literature \citep{masters_2012,cervantessodi_2017}. There are few strong bars in the blue region. In contrast, the weak bar fraction decreases with redder colours. It is interesting to note the similarity between the shapes of the `combined' fraction and Figure 3 in \citet{masters_2011}, a previous GZ study looking at bar fraction against (g-r) colour, especially at redder colours.

The top right plot shows the stellar mass distribution of the different bar types. Again, we see that the strong bar fraction rises with stellar mass. However, this is again not the case for weak bars. Notice the high weak bar fraction for low-mass ($M_{\rm *}$ < 10$^{10} M_{\odot}$) galaxies. \citet{elmegreen_1985} also found more weak bars among low-mass galaxies.

The bottom left panel displays the total SFR. The `combined' bar fraction decreases slightly with total SFR. It is also worth noting the difference between the bar fraction and the histogram. Whereas the strong bar fraction is highest at the lowest total SFRs, it seems that most strong bars (similar to most of the galaxies in our sample) are actually situated in the middle total SFR range (SFR $\sim$ 1 M$_{\odot}$ yr$^{-1}$). 

When looking at the fibre SFRs (bottom right), we see that the strong bar fraction is highest at the lowest and highest fibre SFRs. The `combined' and weak bar fractions display more complicated behaviour with no clear trend.

In conclusion, the general consensus that bars appear more often in massive, red and quiescent galaxies seems to hold true, but only for strong bars. The fraction of weak bars, in contrast, is higher among low-mass and blue galaxies than in massive and red galaxies. 
 
\subsubsection{Colour-magnitude diagram and SFR-mass plane}
\label{subsubsection:colmag_sfrmass}

\begin{figure}
	\includegraphics[width=\columnwidth]{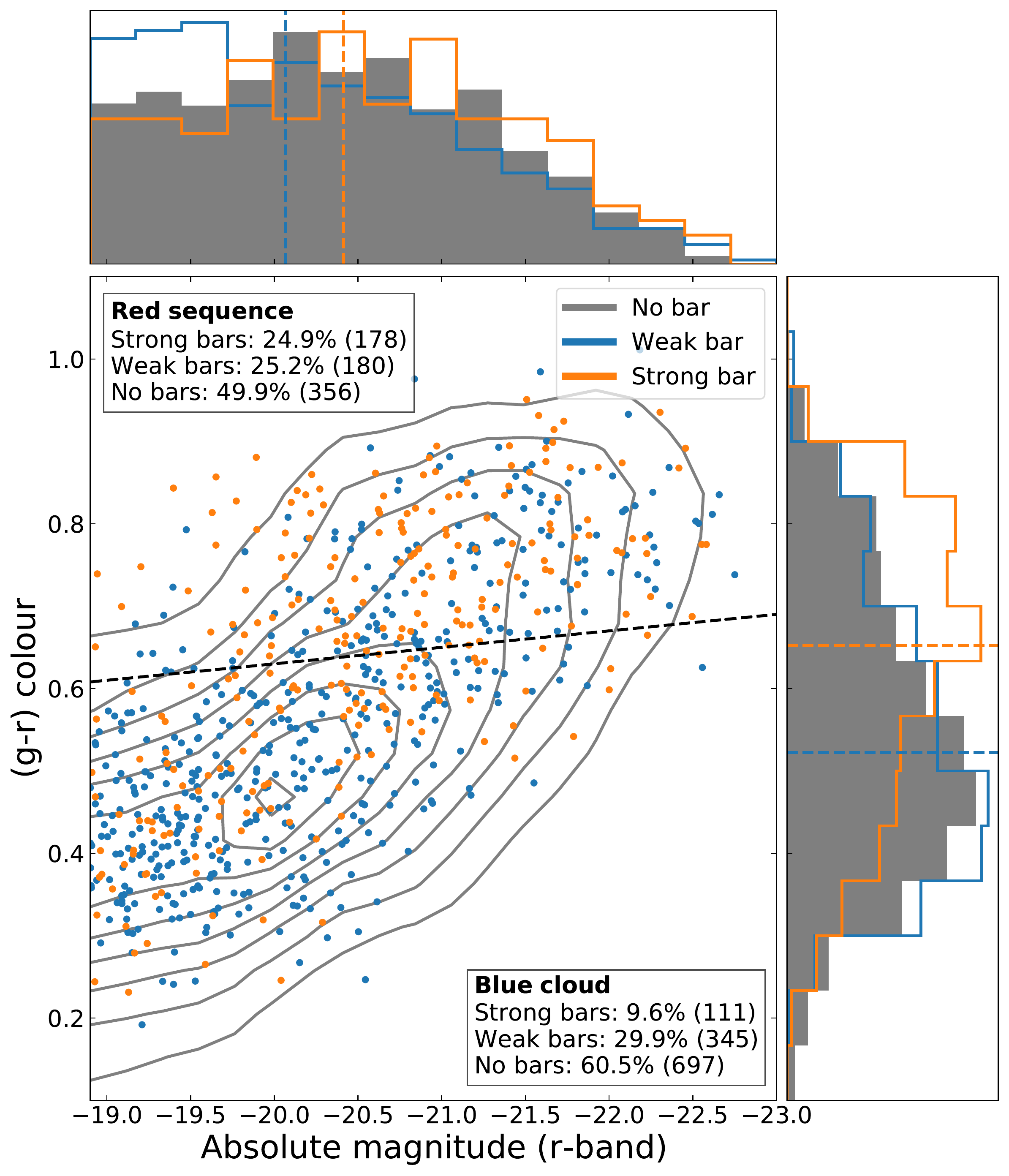}
    \caption{Location of all galaxies with a strong bar (orange) and a weak bar (blue) on the colour-magnitude diagram. The contour plot in the background shows the location of all the galaxies without bars. The dashed line across the plane defines the ``blue edge of the red sequence'' and effectively divides the sample into the blue cloud (underneath the line) and the red sequence (above the line, defined in \citealt{masters_2010}). The histograms on top and on the right side of the main panel are normalised histograms of the absolute r-band magnitude and (g-r) color, respectively, for our three different groups. The galaxies with a weak bar seem to be distributed equally across the plane, while the galaxies with a strong bar are clustered around the red sequence.}
    \label{fig:colmag}
\end{figure}

To expand on the idea that strong bars are preferentially found in massive, red, quiescent galaxies, we study the location of each type of bar on the colour-magnitude plane and SFR-mass plane. The colour-magnitude plane is shown in Figure \ref{fig:colmag}. We used the `blue edge of the red sequence' defined in \citet{masters_2010} to divide our sample into red sequence and blue cloud galaxies.

The weak bar fraction is slightly higher in blue cloud galaxies than in red sequence galaxies, but not by much (\highlight{29.9\%} and \highlight{25.2\%}, respectively). The strong bar fraction is clearly higher in the red sequence (\highlight{24.9\%}) than in the blue cloud (\highlight{9.6\%}). A KS test reveals that the colour distribution of galaxies with strong bars is very significantly different from the unbarred colour distribution (p-value < \highlight{$10^{-15}$}), suggesting that a strong bar influences the colour of its host galaxy and possibly, its evolution. The colour distribution of galaxies with weak bars compared to that of unbarred galaxies is less significantly different (p-value = \highlight{0.012}), but still remarkable.

Interestingly, the absolute magnitude distributions of galaxies with a weak bar and no bar are significantly different (p-value = \highlight{0.009}), which suggests a significant population of low-mass galaxies with a weak bar, similar to what we saw in Section \ref{subsubsection:prop_gals}. This difference is not observed for between galaxies with a strong bar and no bar (p-value = \highlight{0.98}).

\begin{figure}
	\includegraphics[width=\columnwidth]{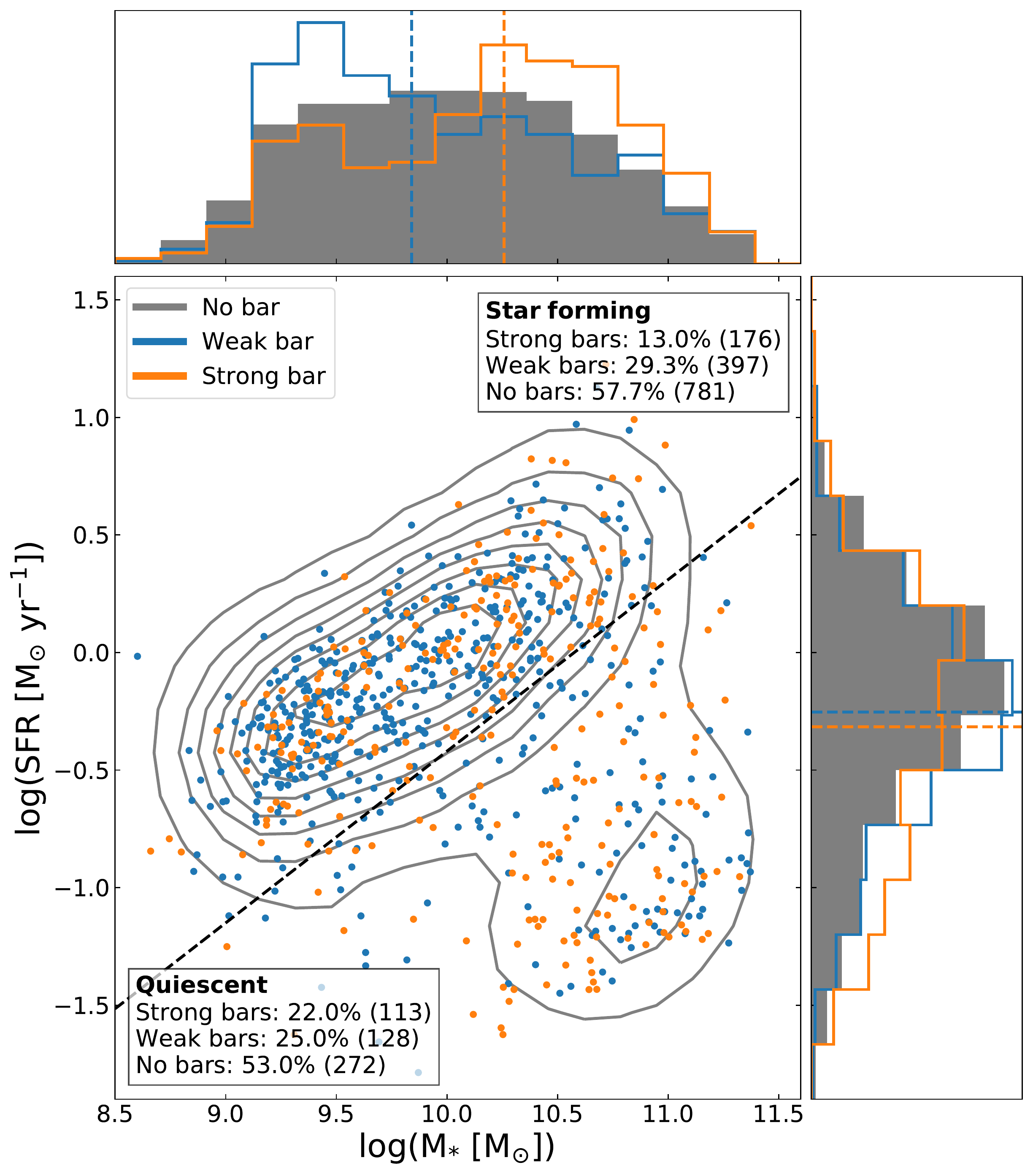}
    \caption{Location of all galaxies with a strong bar (orange) and a weak bar (blue) on the SFR - stellar mass plane. The contour plot in the background shows the location of all the galaxies without bars. The dashed line across the plane  divides the sample into a SF region (above the line) and a quiescent region (under the line, defined in \citealt{belfiore_2018}). The histograms on top and on the right side of the main panel are normalised histograms of the stellar mass and SFR, respectively, for our three different groups. The strong bar fraction is higher in the quiescent region than in the SF region, whereas the weak bar fraction is similar in both regions.}
    \label{fig:sfr_mass}
\end{figure}

We can also plot our galaxies on a SFR - stellar mass plane, shown in Figure \ref{fig:sfr_mass}. On this plane, galaxies tend to cluster naturally into a star forming (SF) group and a quiescent group. Those groups are separated here by using the star formation sequence (SFS) defined by \citet{belfiore_2018}:

\begin{equation}
    \log{ \left( \textrm{SFR}/ \textrm{M}_{\odot}\,\textrm{yr}^{-1} \right) } = (0.73\,\pm\,0.03) \log{ \left(\textrm{M}_{\textrm{*}}/\textrm{M}_{\odot} \right)} - (7.33\,\pm\,0.29) \;,
	\label{eq:sfr_mass_split}
\end{equation}

and assuming anything that is 1$\sigma$ ($=0.39$ dex) below this line is quiescent and everything else star forming (visualised in Figure \ref{fig:sfr_mass} by the dashed line). 

Again, the weak bar fractions are similar in the quiescent group and the SF group (\highlight{25.0\% and 29.3\%}, respectively) whereas the strong bar fraction is higher in the quiescent group than in the SF group (\highlight{22.0\% and 13.0\%}, respectively). However, it is interesting to note that in absolute numbers, there are actually more SF strong bars (\highlight{176}) than quiescent strong bars (\highlight{113}). Also, we can clearly see the difference in the stellar mass distributions between galaxies with a weak bar and unbarred galaxies, (p-value = 0.001). This confirms what we observed above; that there is a significant population of low-mass galaxies with a weak bar. In contrast, there seem to be more high-mass galaxies with a strong bar than unbarred galaxies (p-value < \highlight{$10^{-6}$}).

These results agree with our previous statement: that galaxies with a strong bar drive the observation that bars appear more often in massive, red quiescent galaxies. Weak bars appear in similar fractions in SF and quiescent groups. However, as noted by previous studies \citep{masters_2012}, this can be interpreted in various ways: is it the strong bars that cause the transition from the SF group to the quiescent group by quenching the galaxy or is it easier to form strong bars in quenched galaxies? We discuss this further in Section \ref{subsection:causality}.

\subsection{Bar length}
\label{subsection:barlengths}
\begin{figure*}
    \includegraphics[width = \textwidth]{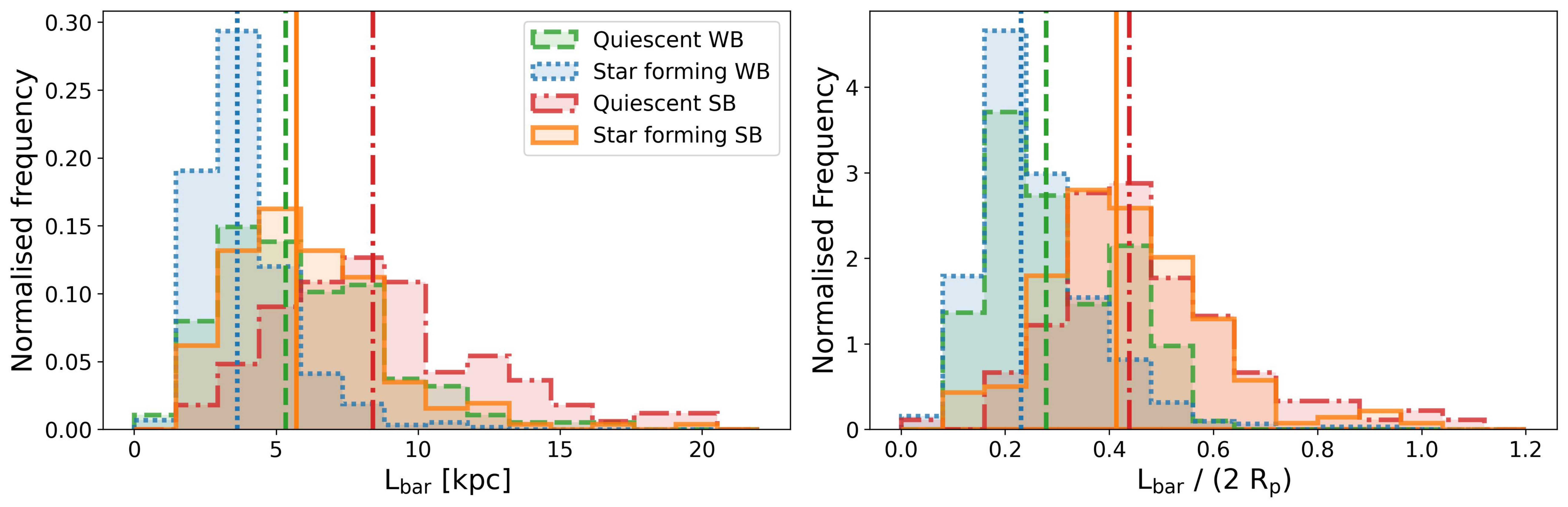}
    \caption{\textbf{Left}: The distribution of bar lengths for the different bar types in absolute units (kpc). \textbf{Right}: The distribution of bar lengths in relative units, obtained by dividing the absolute bar length by the SDSS r-band Petrosian diameter (or twice the Petrosian radius, R$_{\rm p}$). In terms of absolute bar length, quiescent galaxies with a weak (strong) bar are longer than SF galaxies with a weak (strong) bar. When looking at relative bar length, quiescent and SF galaxies of the same bar type have similar bar lengths. A strong bar will span, on average, \highlight{40-45}\% of the Petrosian diameter of its host, while a weak bar will cover \highlight{20-30}\%.}
    \label{fig:barlen}
\end{figure*}

From a quick glance at Figure \ref{fig:SBWB}, we can already see that strong bars tend to be longer than weak bars. This is very unsurprising, as we tell the volunteers that strong bars ``\textit{extend across a large fraction of the galaxy}'' while weak bars are ``\textit{smaller and fainter relative to the galaxy}''. This is quantitatively verified in Figure \ref{fig:barlen}, where we show the distribution of bar lengths of our sample (see Section \ref{subsection:barlen_measurements} for details on how the bar length was measured). We split up the sample, not only in weak and strong, but also into quiescent and SF, according to the cutoff described in Section \ref{subsubsection:colmag_sfrmass}, so we have a total of 4 subsamples. The bar length is shown in kpc in the left panel. Interestingly, we observe that SF and quiescent bars of the same bar type are not the same length. Bars in quiescent galaxies are, on average, longer than the bars in SF galaxies, for both weak and strong bars. Also, the average quiescent weak bar is about as long as the average SF strong bar. The shortest bars are typically weak bars in SF galaxies, while the longest bars are typically strong bars in quiescent galaxies. 

To estimate the length of the bar in relation to its host, we also calculated the relative bar length (right panel of Figure \ref{fig:barlen}), obtained by dividing the absolute bar length by the SDSS r-band Petrosian diameter (the Petrosian radius times two). We find that bars in SF and quiescent galaxies with the same bar type have roughly the same relative length. Strong bars are on average longer than weak bars in terms of relative bar length. A weak bar will, on average, cover \highlight{20-30}\% of the Petrosian diameter, whereas a strong bar will span over, on average, \highlight{40-45}\% of the Petrosian diameter.

Again, it is worth stressing that we expected strong bars to be longer than weak bars, as this is what we asked the volunteers to select for. However, it is interesting to note that there is still quite an overlap in (absolute and relative) bar length between weak and strong bars. 

Interestingly, we see that strong bars in quiescent galaxies are longer (in terms of absolute bar length) than strong bars in SF galaxies. A possibility is that bars are still growing (and possibly helping to quench their host) when in SF galaxies and are fully grown once their host becomes quiescent. Additionally, the relative bar size of strong bars in quiescent and SF galaxies is similar, which suggests that the bar co-evolves with the entire galaxy and showcases that bars have an important role in galaxy evolution. Similar results are observed for weak bars.

However, the plot on the left side of Figure \ref{fig:barlen} can be a natural consequence of the volunteers classifying bars into weak and strong based on the relative bar length and the fact that quiescent galaxies tend to be larger than SF galaxies. Please refer to Appendix \ref{app:coevolution} for a more detailed discussion on this topic. 

\subsection{Addressing causality}
\label{subsection:causality}

We have found that the fraction of galaxies with a strong bar is notably higher in quiescent galaxies than in SF galaxies, whereas this was not found for weak galaxies. However, it is not clear whether strong bars help to quench their host or whether it is easier to form strong bars in quenched galaxies.

In an attempt to address this problem of causality, we look more closely at the SFR in the central 3 arcsec of the galaxy - the fibre SFR - as that is where we expect to see a difference in SFR caused by the bar if bars drive gas to the centre. Additionally, the global SFRs estimates from \citet{brinchmann_2004} are in part based on the colour outside the fibre. However, we know from \citet{kruk_2018} that bars are redder than discs. Given that we are comparing different bar types, we are unsure if this will affect our bar types differently and therefore decided to work solely with SFR estimates based on emission lines in the fibre instead.

As the Kennicutt-Schmidt law \citep{kennicutt_1998, kennicutt_1998b} implies a relation between SFR and gas, we are also interested in potential differences in gas mass as well. For this, we use the gas mass measurements from the ALFALFA catalog of extragalactic HI sources \citep{giovanelli_2005,haynes_2011,haynes_2018}. In order to accommodate for the non-detections in ALFALFA (see Section \ref{subsection:alfalfa}), we perform a survival analysis on the data using the Python package `lifelines' \citep{lifelines} \footnote{lifelines: \url{https://lifelines.readthedocs.io/en/latest/\#}}. Survival analysis is a statistical technique that allows us to compare two populations that have left-censored data points in a certain parameter (in our case: ALFALFA upper limits for gas mass) by correctly constructing the cumulative density function (CDF) for that parameter. We use the well established Cox’s proportional hazards model \citep{cox_1972} to quantify the effect of every bar type on the CDF. It is a type of regression that allows for censored data and is often used in conjunction with survival analysis. Finally, we also divide our sample in a quiescent and SF group, as shown in Figure \ref{fig:sfr_mass}, to compare like with like: SF strong bars with SF weak bars and quiescent strong bars with quiescent weak bars.

\begin{figure*}
	\includegraphics[width=\textwidth]{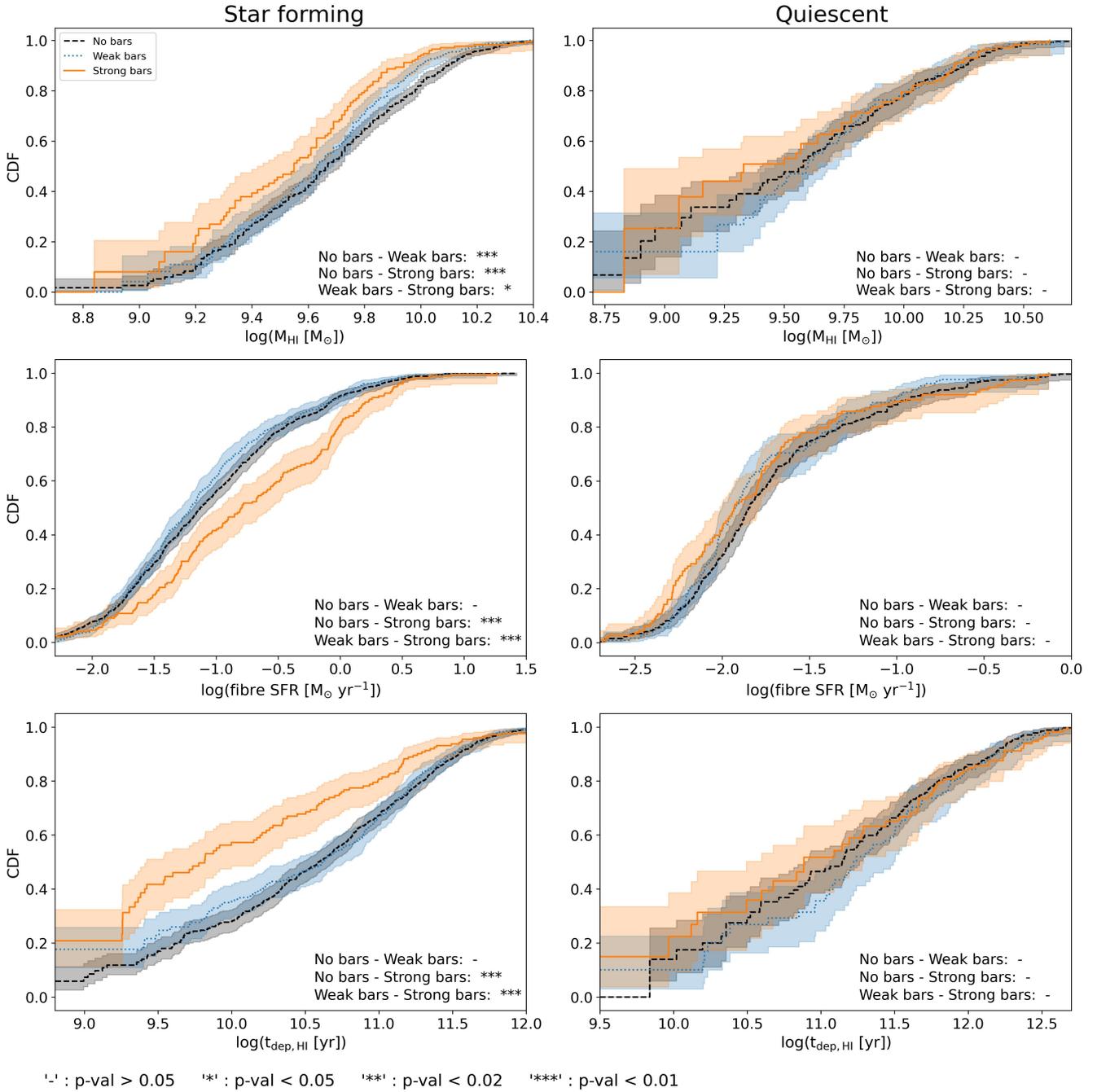}
    \caption{The results from a survival analysis on galaxies with no bar (black dashed), a weak bar (blue dotted) and a strong bar (orange full) in the SF group (left) and quiescent group (right). The top row displays the results for the HI gas mass, the middle row for fibre SFR and the depletion timescale is shown in the bottom row. The results are shown in the form of cumulative density functions (CDFs). The results of the Cox's proportional hazards model are displayed in the bottom-right corner of every panel.  If the p-value is below 0.01 (i.e., they are distinct), it is marked with `***'. If the p-value was between 0.01 and 0.02, it is denoted with `**'. If it was between 0.02 and 0.05, `*' was used. If it is above 0.05, it was denoted with `-'. The SF galaxies with a strong bar are statistically significantly different from SF galaxies with a weak bar or no bar, especially in terms of fibre SFR and depletion timescale. Strong bars in SF galaxies have the highest fibre SFRs and shortest depletion timescales. There are no significant differences among the bar types in the quiescent galaxies.}
    \label{fig:surv_comb}
\end{figure*}

The results are shown in Figure \ref{fig:surv_comb}. The top row displays the results for HI gas mass. It seems that SF galaxies with a strong bar or a weak bar have significantly less HI gas than SF unbarred galaxies, with galaxies with a strong bar having the lowest amount of HI gas. The CDFs of the fibre SFRs are shown in the middle row of Figure \ref{fig:surv_comb}. The SF galaxies with a strong bar have significantly higher values for fibre SFR than both SF galaxies with a weak bar or no bar. There are no significant differences between SF galaxies with a weak bar and unbarred galaxies in terms of fibre SFR. No significant differences were observed among quiescent galaxies.

We can put a timescale on these effects by combing the two parameters above (fibre SFR and HI gas mass) into the depletion timescale:

\begin{equation}
    t_{\textrm{dep}}=\frac{\rm{M}_{\textrm{gas}}}{\textrm{fibre SFR}}\;,
	\label{eq:tdep}
\end{equation}

which probes how quickly a galaxy will consume its gas reservoir for a given SFR \citep{janowiecki_2020}. Galaxies with a long depletion timescale will take a long time before they run through their gas reservoir, and the opposite is true for galaxies with short depletion timescales (assuming all the gas in the reservoir will be available for star formation). The depletion timescale might be a too simplistic metric, as \citet{janowiecki_2020} noted that it assumes a constant SFR and that gas recycling is non-existent, while other work \citep{kannappan_2013} has shown that refueling on $\sim$Gyr timescales can happen in SF galaxies. Also, we use fibre SFR to calculate the depletion timescale, while others use the global SFR. Thus, the depletion timescales shown here are in fact depletion timescales for the centre of the galaxy (or we have to assume little to no SF outside the fibre, which is unlikely). Additionally, we have to point out that, while the fibre SFR only probes the centre, the gas mass estimates are global gas masses. Finally, we are also using HI gas measurements rather than molecular gas. All this results in the relatively high values for the depletion timescale. However, these issues affect all galaxies in our sample in the same way, so it is still a useful metric.

The depletion timescale is shown in the bottom row of Figure \ref{fig:surv_comb}. We see that the SF galaxies with a strong bar have a significantly shorter depletion timescale than the SF galaxies with a weak bar or no bar. This result implies that SF galaxies with a strong bar are rapidly evolving and use up their gas quicker due to increased central SFRs. However, it is important to note that this was only seen in SF galaxies. The various bar types show almost no significant differences when their host is quiescent.

However, stellar mass also correlates with both gas mass and (fibre) SFR \citep{brinchmann_2000,brinchmann_2004,noeske_2007,lara-lopez_2010}. In addition, as we have noted above, there are differences in the stellar mass distribution between galaxies with a weak and strong bar.

This bias is filtered out by plotting the parameters above against stellar mass and assessing their relation with the Python package `linmix' \footnote{linmix: \url{http://linmix.readthedocs.org/}}, which is based on the hierarchical Bayesian model of \citet{kelly_2007}. This technique allows us to fit a linear relationship to data with two-dimensional uncertainties as well as correctly taking into account the censored data from ALFALFA. It works by running a Markov chain Monte Carlo (MCMC) algorithm to sample the posterior distribution, given the data, which one can use to correctly estimate the errors on the fit. As we only observe significant differences between the bar types in the SF galaxies in Figure \ref{fig:surv_comb}, we only look at the SF galaxies here.

\begin{figure*}
	\includegraphics[width=\textwidth]{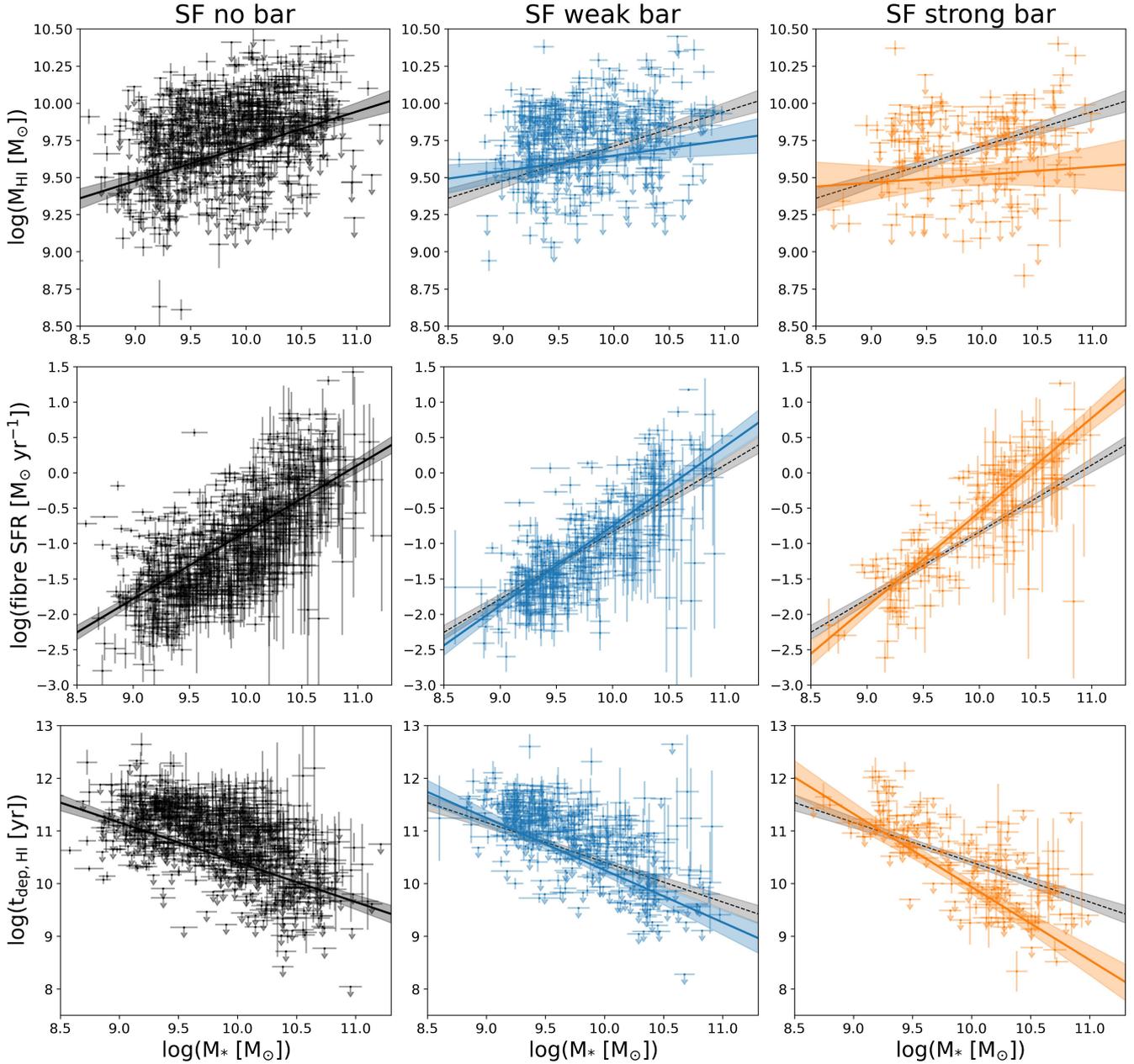}
    \caption{
    One-dimensional trends between HI gas mass (top row), fibre SFR (middle row) and depletion timescale (bottom row) and stellar mass. The trends were assessed using the linmix Python package, which is based on the hierarchical Bayesian model of \citet{kelly_2007}. The fits were created separately for SF galaxies with no bar (left column), SF galaxies with a weak bar (middle column) and SF galaxies with a strong bar (right column). The trends of the unbarred galaxies are drawn over the other panels as well (in black, dashed) to facilitate comparison. The 2$\sigma$ errors on the fit are displayed with the light shading. The trends of the SF galaxies with a weak bar seem very similar to the trends of the SF unbarred galaxies. In contrast, the trends of the SF galaxies with a strong bar are very different from that of the SF unbarred galaxies: the SF galaxies with a strong bar have lower gas mass, high fibre SFRs and shorter depletion timescales than SF unbarred galaxies, except at the lowest stellar masses.} 
    \label{fig:linmix_comb}
\end{figure*}

The one-dimensional relations created with linmix are visualised in Figure \ref{fig:linmix_comb}. The top row shows HI gas mass for unbarred galaxies (left column, black), galaxies with a weak bar (middle column, blue) and galaxies with a strong bar (right column, orange). The contours show the 2$\sigma$ uncertainty on the fit. To facilitate comparison, the trend for the unbarred galaxies is displayed with a dashed line in all subplots. We see that galaxies with a strong bar have lower HI gas masses, except at the lowest stellar mass range (10$^{9.5} M_{\odot}$ < $M_{\rm *}$) compared to unbarred galaxies. The trend for galaxies with a weak bar is also lower and less steep than the trend for unbarred galaxies, but the 2$\sigma$ uncertainty contours partially overlap, suggesting that this isn't as significantly different.

The middle row of Figure \ref{fig:linmix_comb} displays the trends for fibre SFR. Galaxies with a weak bar and unbarred galaxies do not have significantly different trends. However, the trend of galaxies with a strong bar is steeper than that of the unbarred galaxies, which shows that SF galaxies with a strong bar have higher fibre SFRs, especially at higher stellar masses. The depletion timescale trends show a similar result: galaxies with a weak bar and unbarred galaxies do not differ much, while galaxies with a strong bar have shorter depletion timescales, especially at higher stellar masses.

We have also made Figures \ref{fig:surv_comb} and \ref{fig:linmix_comb} for our sample with a more stringent inclination cut of $i < 45 ^{\circ}$ (not shown). The inclination measurements used were calculated using the adaptive moments from SDSS DR16 \citep{ahumada_2020}. We saw that the difference between strong and weak bars became more prominent and more statistically different for fibre SFR and depletion timescale. This is most likely because a stricter inclination threshold results in cleaner samples (at the cost of a lower sample size). However, the opposite was true for HI gas mass, whose trends became more uncertain and less statistically different. We assume this is due to the extra uncertainties associated with the non-detections in ALFALFA and the resultant few actual detections when limiting our sample to $i < 45 ^{\circ}$. To avoid this low sample size, we decided to not apply this additional inclination threshold throughout the paper and used our current threshold based on $p_{\textrm{not edge-on}}$ instead.

The two different methods applied to our sample show the same result: that SF galaxies with a strong bar differ significantly from SF galaxies with a weak bar and unbarred galaxies, who do not differ from each other as much. SF galaxies with a strong bar have lower HI gas mass, possibly due to higher fibre SFRs, which results in shorter depletion timescales. This implies that SF galaxies with a strong bar are rapidly evolving galaxies, which quench faster than their weakly barred and unbarred counterparts.

\subsection{Are weak and strong bars distinct populations?}
\label{subsection:weak_strong_distinct_pop}
\begin{figure*}
	\includegraphics[width=\textwidth]{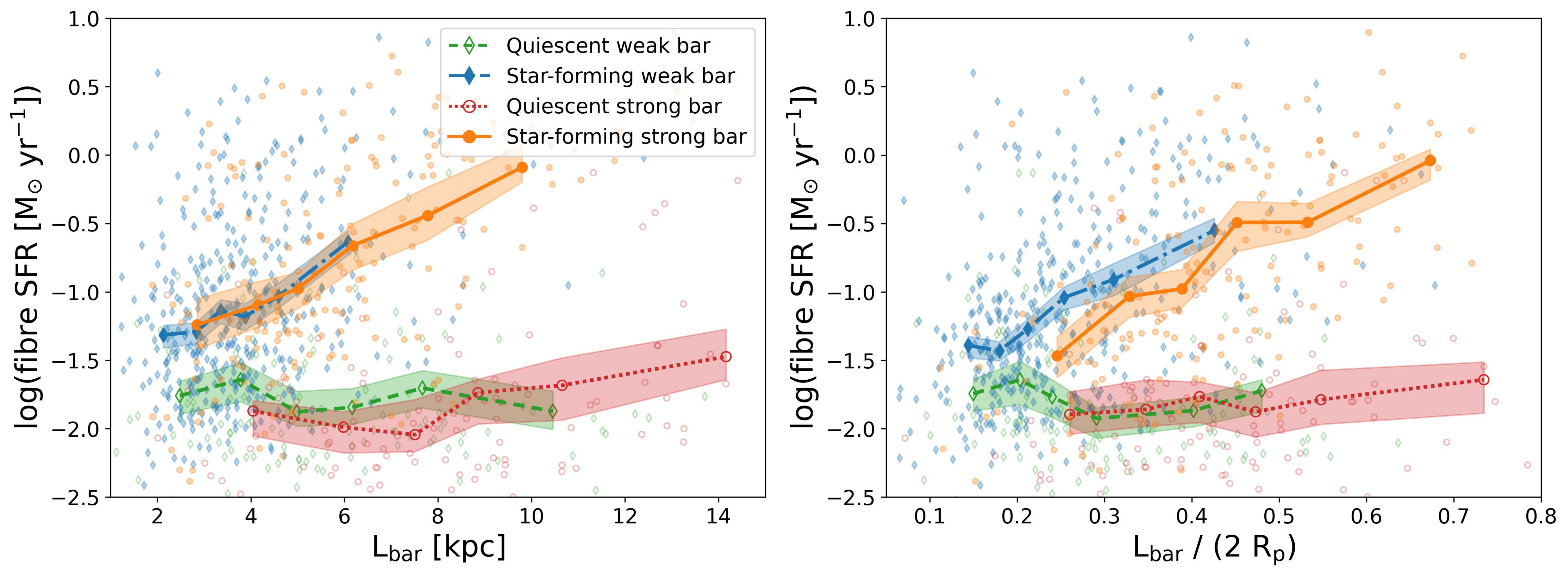}
    \caption{\textbf{Left}: The absolute bar length (in kpc) against fibre SFR for quiescent galaxies with a weak bar, quiescent galaxies with a strong bar, SF galaxies with a weak bar and SF galaxies with a strong bar. \textbf{Right}: The relative bar length (obtained by dividing the absolute length by the SDSS r-band Petrosian diameter) against fibre SFR. Please refer to Figure \ref{fig:barlen} for histograms of the bar lengths. At a given bar length (absolute or relative), the fibre SFR for SF galaxies with a strong bar is not higher than the fibre SFR for SF galaxies with a weak bar, contrary to what was was observed in Figures \ref{fig:surv_comb} and \ref{fig:linmix_comb}. The contours represent the 3$\sigma$ region after bootstrapping the data 10,000 times and retaining 90\% of the data for each iteration.} 
    \label{fig:fibresfr_barsize}
\end{figure*}

We have shown that, on average, SF galaxies with a strong bar have higher fibre SFRs than SF galaxies with a weak bar at a given stellar mass. The next question to ask is: is there a measurable difference between strong and weak bars when controlling for bar length? In Figure \ref{fig:fibresfr_barsize}, we plot fibre SFR against the length of the bar for four different subgroups: SF galaxies with a strong bar, SF galaxies with a weak bar, quiescent galaxies with a strong bar and quiescent galaxies with a weak bar. The length of the bar is measured in absolute units (left panel) and relative  units, obtained by dividing the absolute length by the Petrosian diameter (right panel).  Interestingly, with both measures of bar length, we see that galaxies with a strong bar do not have higher fibre SFR than galaxies with a weak bar for a given bar length. This suggests that both strong and weak bars have similar influences on their host galaxy (i.e., funnel gas to the centre, where it is used to increase SFR) when controlling for bar length. However, for SF galaxies in general, we do see that fibre SFR positively correlates with bar length (absolute and relative), suggesting that the strongest and longest bars have the highest fibre SFR and affect their host the most. This is not the case for the quiescent galaxies.

This leads us to ask a more fundamental question: are strong and weak bars fundamentally distinct physical phenomena and distinct populations? Or are they just two labels we give to bars that are on opposite ends of a continuum of bar types? If strong and weak bars are distinct phenomena, then we expect to see differences between the two bar types when controlling for bar length. The results of Figure \ref{fig:fibresfr_barsize} give weight to the latter option, as it shows that, for a given bar length (absolute or relative), there is no difference in fibre SFR. 

\begin{figure*}
	\includegraphics[width=\textwidth]{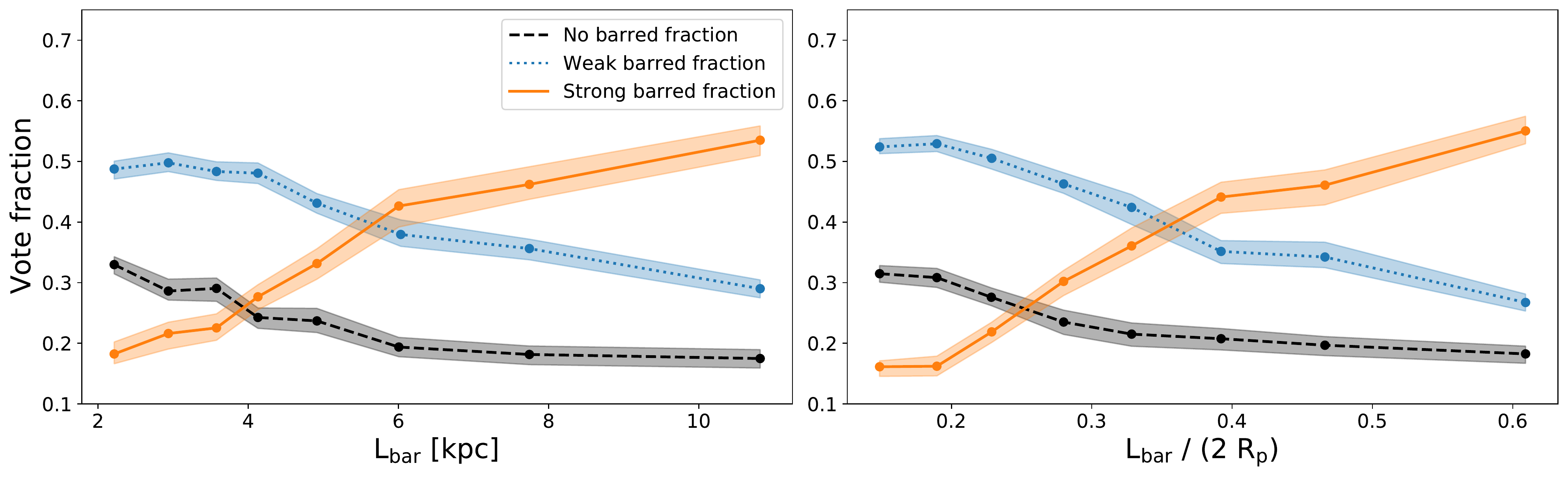}
    \caption{\textbf{Left}: The absolute bar length (in kpc) against the average strong, weak and unbarred vote fraction. \textbf{Right}: The relative bar length (obtained by dividing the absolute length by the SDSS r-band Petrosian diameter). Volunteers become increasingly confident that a bar is strong if it is longer. The contours represent the 3$\sigma$ region after bootstrapping the data 10,000 times and retaining 90\% of the data for each iteration.} 
    \label{fig:pbar_lbar}
\end{figure*}

We can see that the GZD bar vote fractions correctly reflect the continuous nature of bar types. In Figure \ref{fig:pbar_lbar}, we plot the vote fractions ($p_{\rm no \; bar}$, $p_{\rm weak \; bar}$ and $p_{\rm strong \; bar}$) against bar length. These results show that, as the bar becomes longer (both absolute and relative length), volunteers are more likely to vote that the bar is strong. This is not surprising, as we tell the volunteers that strong bars are longer. However, this figure also shows that volunteers become increasingly confident that a bar is strong if it is longer. After a bar length of $\sim$\highlight{6} kpc (or a relative bar length of $\sim$\highlight{35}\%), we start to see more strong than weak bars. However, there is no distinctive bar length cut-off before (after) which a bar is always weak (strong). This emphasizes the continuous nature of bar types, rather than the dichotomy that separates weak and strong bars in distinct separate classes.

\begin{figure*}
	\includegraphics[width=\textwidth]{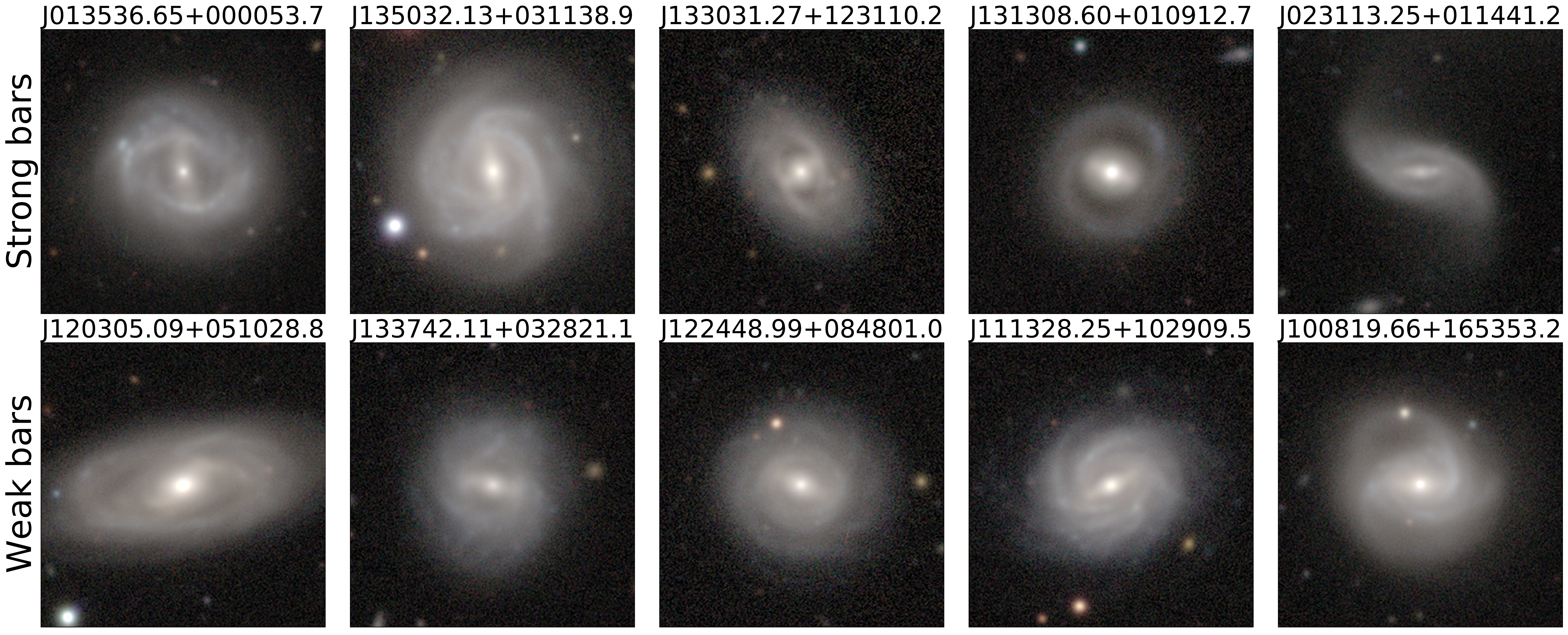}
    \caption{DECaLS postage stamps (59x59 arcsec) of galaxies with intermediate relative bar lengths (\highlight{0.27} < L$_{\rm bar}$/Petrorad < \highlight{0.37}) and redshifts between 0.027 < $z$ < 0.035. Galaxies classified by GZD as having a strong bar are shown in the top row, while galaxies with a weak bar are in the bottom row. About \highlight{23\%} of the bars in our sample fall within this intermediate relative bar length range. The authors of this paper find it hard to distinguish between the weak and strong bars presented here, especially considering the obvious morphological differences between weak and strong bars in Figure \ref{fig:SBWB}.}
    \label{fig:bars_notsure}
\end{figure*}

In Figure \ref{fig:bars_notsure}, we show postage stamps of randomly selected galaxies in the intermediate relative bar length (L$_{\rm bar}$/Petrorad) region where there are both strong and weak bars (\highlight{0.27} < L$_{\rm bar}$/Petrorad < \highlight{0.37}). The top row shows bars that are classified as strong bars by GZD, while the bottom row shows weak bars. Around \highlight{23\%} (\highlight{188 out of 814}) of the barred galaxies in our sample fall within this intermediate relative bar length range. After visual inspection of those galaxies, it seems clear that morphologically speaking, galaxies with strong and weak bars of intermediate relative length look very similar to each other, unlike the obviously strong and weak bars shown previously in Figure \ref{fig:SBWB}.

These results show that at an intermediate relative bar length (absolute or relative), strong and weak bars seem to be hard to differentiate. This suggests that strong and weak bars are not distinct physical phenomena and hints at the continuous nature of bar types.

\section{Discussion}
\label{section:discussion}

\subsection{Difference in bar length between strong and weak bars}

In Figure \ref{fig:barlen} in Section \ref{subsection:barlengths}, we compared the absolute bar length and relative bar length for weak and strong bars, separated into SF and quiescent subsamples. We see that strong bars are typically longer than weak bars (both in terms of absolute and relative size). This is expected, as this corresponds to the guidelines given to the volunteers while classifying. However, we also found that, when measuring the bar length in kpc (absolute size) quiescent galaxies have longer bars, regardless of bar type. Bars in quiescent galaxies are, on average, longer than bars in SF galaxies, for both weak and strong bars. This can be explained by supposing that bars in SF galaxies have not reached their full size yet and are still growing, whereas bars in quiescent galaxies have reached their maximum size and cannot grow anymore.

This difference disappears when measuring the bar size relative to the Petrosian diameter. Strong bars typically cover \highlight{40-45}\% of the Petrosian diameter, whereas weak bars only cover around \highlight{20-30}\%, regardless whether the host is SF or quiescent. This observation hints that bars do not evolve independently from their host. 

However, it is possible that the observed results are a simple consequence of quiescent galaxies being bigger than SF galaxies and that volunteers classify bars into weak or strong based on the relative bar length (as we expect them to). This possibility is discussed in more detail in Appendix \ref{app:coevolution}. While the interpretation is open to debate, we can conclude that in terms of absolute bar length, bars in quiescent galaxies are longer than bars in SF galaxies, irrespective of bar type. In addition, we found that in terms of relative bar length, strong bars cover \highlight{40-45\%} of the Petrosian diameter, while weak bars cover \highlight{20-30\%}, irrespective of whether the host galaxy is quiescent or SF.

\subsection{Strong bars are more common among quiescent galaxies}

In Figure \ref{fig:barfrac_plots} in Section \ref{subsubsection:prop_gals}, we looked at the properties of the galaxies hosting bars. We showed that the strong bar fraction increases monotonically with (g-r) colour and stellar mass. The strong bar fraction is also highest at lower SFR bins, although there is a significant population of strong bars found at higher SFR (and fibre SFRs). This is in agreement with the literature, as many other studies have shown that bars are more likely to be found in galaxies that have higher stellar mass, less gas, redder colours and low SFRs \citep{nair_2010b, masters_2012,vera_2016,cervantessodi_2017}. However, we have only observed these results for strong bars, not weak bars. 

\citet{erwin_2018} showed that, in a sample drawn from the Spitzer Survey of Stellar Structure in Galaxies (S4G), the bar fraction is constant over a range of (g-r) colours and gas fractions. Their bar fraction does not increase, but rather decreases for stellar masses higher than $\sim$ 10$^{9.7} M_{\odot}$. These results are in contrast to many SDSS-based studies cited above. \citet{erwin_2018} argues that this apparent contradiction can be explained if SDSS-based studies miss bars in low-mass blue galaxies. In Figures \ref{fig:gzd_gz2_comb} and \ref{fig:barfrac_plots}, we showed that the newly detected bars in GZD (compared to GZ2) are weak bars in low-mass blue galaxies. Nevertheless, the `combined' bar fraction in Figure \ref{fig:barfrac_plots} is not constant over (g-r) colour and agrees well with \citet{masters_2011} for redder colours ((g-r) colour > 0.5). Additionally, our `combined' bar fraction remains roughly constant over stellar mass. As mentioned before, we conclude that strong bars drive the trends of bar fraction with (g-r) colour, stellar mass and SFR observed in other studies \citep{nair_2010b, masters_2011, masters_2012,vera_2016,cervantessodi_2017}. However, the addition of weak bars in low-mass blue galaxies is insufficient to resolve the apparent disagreement between \citet{erwin_2018} and many SDSS-based studies \citep{masters_2011,masters_2012,vera_2016,cervantessodi_2017,kruk_2018}, which instead seems likely to be due to the very different sample selection of the S4G and SDSS galaxy samples. For example, the median stellar mass of the sample used in \citet{erwin_2018} is $\sim 10^{9.6} M_{\odot}$ (based on their Figure 4 and the bins in the top-left panel of their Figure 5). However, the median stellar mass of our sample is $10^{10.6} M_{\odot}$. As stellar mass correlates with many parameters (including bar length), this can have major consequences. Additionally, as \citet{erwin_2018} notes,  there is also the issue of resolution to consider. With an r-band FWHM of 1.18" from DECaLS \citep{dey_2019} and a mean redshift of 0.036, the mean linear resolution of our sample is approximately 834 pc, which is higher than the 165 pc of \citet{erwin_2018}. This explains why they observe many sub-kpc bars, while we do not. These differences in stellar mass and resolution will manifest themselves in the conclusions, so a more detailed analysis is needed for a proper comparison with \citet{erwin_2018}.

In Figures \ref{fig:colmag} and \ref{fig:sfr_mass} in Section \ref{subsubsection:colmag_sfrmass}, we observe higher strong bar fractions in the red sequence and in quiescent galaxies (\highlight{24.9\% and 22.0\%}, respectively) than in the blue cloud and in SF galaxies (\highlight{9.6\% and 13.0\%, respectively}). The fraction of galaxies with a weak bar is roughly the same in red sequence and in quiescent galaxies (\highlight{25.2\% and 25.0\%}, respectively) compared to the blue cloud and in SF galaxies (\highlight{29.9\% and 29.3\%, respectively}). This data can be interpreted in two different ways. Either it is easier to form strong bars in quenched galaxies, or only strong bars help to quench the galaxy \citep{masters_2012}.

\subsection{Strong bars in star forming galaxies facilitate quenching}

In an attempt to see whether strong bars help to quench their host or if it is easier to form a strong bar in a quenched galaxy, we performed a detailed analysis on fibre SFR, HI gas mass and the depletion timescale in Figures \ref{fig:surv_comb} and \ref{fig:linmix_comb} in Section \ref{subsection:causality}. We found that, on average, SF galaxies with a strong bar have higher fibre SFRs, lower gas masses and shorter depletion timescales for a given stellar mass, compared to SF unbarred galaxies. This suggests that, on average, SF galaxies with a strong bar are more rapidly evolving galaxies and are more efficient in star formation, compared to unbarred galaxies. In contrast, SF galaxies with a weak bar do not differ significantly from galaxies without bars in terms of fibre SFR and depletion timescale. Similarly, no significant differences were observed between quiescent galaxies with a strong, weak or no bar. Do note that we are describing average properties and that there is scatter around these trends. A particular weakly barred galaxy can still have a higher fibre SFR than another galaxy with a strong bar. However, on average, the trends described above apply.

It has been suggested before by various other authors that bars can affect and potentially quench their host through ``secular evolution'' \citep{kormendy_2004,sheth_2005,athanassoula_2007,masters_2011,cheung_2013,kruk_2018,efthymiopoulos_2019}. This can be done by bars funnelling gas inwards \citep{athanassoula_1992, athanassoula_2013,villa-vargas_2010} and once the gas is concentrated in the centre, it is used to increase SFR. Increasing the SFR in a galaxy speeds up the rate of gas consumption, thereby facilitating the transition from star forming to quenched. An increase in SFR in barred galaxies has been observed before \citep{alonso_herrero_2001, hunt_2008, ellison_2011,coelho_2011, hirota_2014,janowiecki_2020,magana_serrano_2020,lin_2020}. However, we conclude that, on average, \textbf{only strong bars in SF galaxies help to quench their host through secular evolution}.

We have observed an increase in fibre SFR for SF galaxies with a strong bar, meaning within the central 3 arcsecs of the galaxy. Further constraining the spatial distribution of the observed increase in SFR is of much interest and could be done with integral field spectroscopic surveys such as Mapping Nearby Galaxies at Apache Point Observatory (MaNGA, \citealt{bundy_2015}). As \citet{gavazzi_2015} note, a bar only helps to quench the inner regions of a galaxy, which gives the bar its usual red colour. A similar phenomena is found in the `star formation desert' by \citet{james_2018} as well as the observed gas-depleted regions in barred galaxies \citet{spinoso_2017, george_2019, newnham_2020}. Thus, other quenching mechanisms, such as environmental quenching, are likely to be needed to work in tandem with bar quenching in order to fully quench the galaxy \citep{smethurst_2017}. 

Nevertheless, some studies have shown that bars do not cause an increase of SFR/SFE \citep{sheth_2000,khoperskov_2018}. \citet{watanabe_2011} found that within the bar radius the SFE is similar to the disk region. Even more, \citet{sheth_2000} found lower SFR in the region between the centre and bar ends. These lower SFR/SFE in strongly barred regions have been explained by strong gas flow and/or shear effects induced by the bar potential, which lowers SFR \citep{athanassoula_1992,reynaud_1998,sheth_2000,sorai_2012, meidt_2013, momose_2010,nimori_2013,krishnarao_2020} or fast cloud-cloud collisions \citep{fujimoto_2014b, maeda_2018, fujimoto_2020,maeda_2021}. In contrast to the studies cited above, we see an increase in fibre SFR. However, the fibre only covers the central 3 arcsec of every galaxy, which corresponds to the central \highlight{0.69 - 2.93} kpc in our sample. As shown in Figure \ref{fig:barlen} in Section \ref{subsection:barlengths}, bars (especially strong bars) are typically larger than that, which means that the fibre only covers the central part of the bar. This suggests that the increase in SFR observed here is indicative of an increase in SFR in the centre of the galaxy, and not in the entire bar or bar-end region, which is where the studies cited above typically look at. Additionally, many of these studies do not differentiate between weak and strong bars, which could also explain the apparent discrepancy, as we have only found an increase in fibre SFR in SF galaxies with a strong bar.

We have provided evidence that SF strong bars help to quench their host. However, the simulations of \citet{athanassoula_2013} found that a bar starts forming later and forms more slowly if the gas fraction is high. At the end of their simulation, the bar was much weaker in gas-rich galaxies. This is a recurring finding \citep{villa-vargas_2010,cervantessodi_2017}. \citet{sheth_2008} has shown that at high redshifts, red disk galaxies obtain their bar earlier than blue spirals. These findings would suggest that it is easier to form bars in quenched galaxies, not that bars facilitate quenching. However, it is important to note that our results do not exclude this possibility. Instead, our results suggest that SF strong bars facilitate quenching, and it may still be true that strong bars are easier to form in quenched galaxies at the same time.

\subsection{Are weak and strong bars different phenomena?}

After having established that, on average, SF galaxies with a strong bar have higher fibre SFRs than SF galaxies with a weak bar for any given stellar mass, we have shown that this is not true when controlling for bar length in Figure \ref{fig:fibresfr_barsize} in Section \ref{subsection:weak_strong_distinct_pop}. In other words, in terms of fibre SFR, weak and strong bars of a fixed bar length are indiscriminable. This compelled us to look at images of weak and strong bars in galaxies that have `intermediate' bar lengths (\highlight{0.27} < L$_{\rm bar}$/Petrorad < \highlight{0.37}) in Figure \ref{fig:bars_notsure} and concluded that they look very similar. 

If weak and strong bars were physically different phenomena, we would expect to see differences in fibre SFR at a fixed bar length. As we did not see any, we have to conclude that weak and strong bars are not separate and distinct barred phenomena. Rather, we propose that all barred galaxies are part of a continuum of bar types, which have previously been divided into two populations (weak and strong). With this continuum in mind, it makes more sense to talk about `stronger' and `weaker' bars, rather than `strong' and `weak' bars.

Even though one of the conclusions of this paper is that weak and strong bars are not fundamentally distinct physical phenomena, we believe it is still worthwhile to characterise and differentiate between the ends of that continuum, as the extremes do have varying effects on their host (e.g. in fibre SFR, gas mass and depletion timescale). This is most easily done by splitting the bar population into two or more categories, as done by GZD (weak and strong) and \citet{nair_2010} (weak, intermediate and strong). One should only be cautious while choosing a threshold and should keep the continuous nature of bar types in mind while interpreting the results. 

\begin{figure*}
	\includegraphics[width=\textwidth]{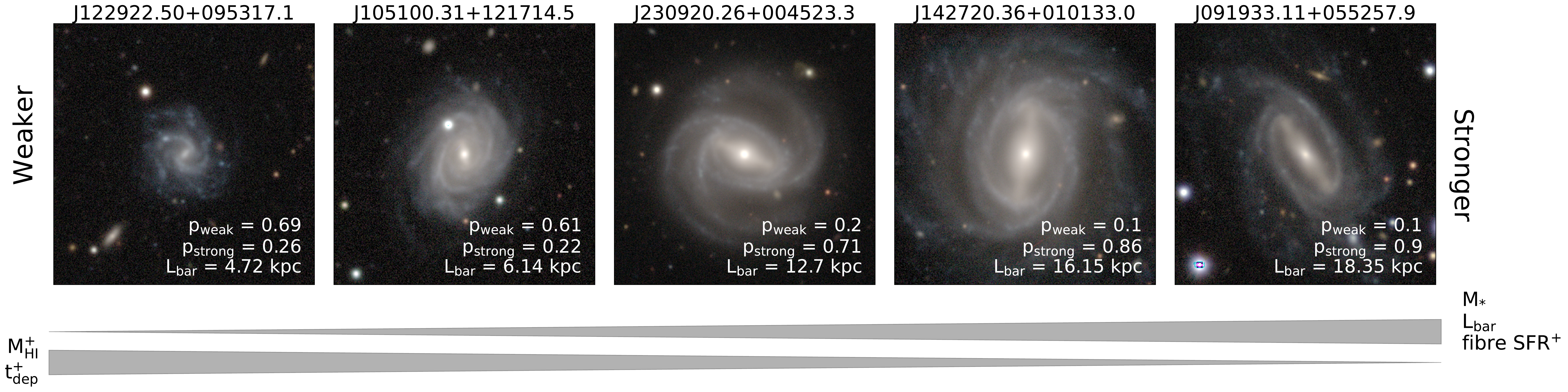}
    \caption{A visualisation of the proposed bar continuum with DECaLS postage stamps (93x93 arcsec). The continuum is shown going from weaker bars (left) to stronger bars (right). Various parameters scale together with this continuum. For example: bar length (absolute and relative), p$_{\rm strong}$, stellar mass and fibre SFR. Conversely, various parameters are inversely proportional to bar strength, such as p$_{\rm weak}$, HI gas mass and depletion timescale. Please note that depletion timescale, HI gas mass and fibre SFR only show these trends if we only consider SF galaxies (denoted by the plus sign in superscript). The bar continuum shown here is based on the absolute bar length (L$_{\rm bar}$), but it can also be based on other parameters, such as relative bar length or a combination of p$_{\rm weak}$ and p$_{\rm strong}$.} 
    \label{fig:barspectrum}
\end{figure*}

The proposed bar continuum is visualised in Figure \ref{fig:barspectrum} using DECaLS postage stamps. The galaxies are ordered from  having weaker bars to stronger bars (from left to right). In this paper, we have shown that various parameters scale with this bar continuum on a population level, such as bar length (absolute and relative) and stellar mass. When we only consider SF galaxies, we have also shown that only the strongest bars increase the fibre SFR, decrease HI gas mass and lower the depletion timescale, which facilitates the quenching process. This is not observed for weaker bars and illustrates how the position of the bar on this continuum can have a big effect on the host galaxy.

Multiple studies have found that bars can grow longer and stronger over time \citep{athanassoula_2003, kim_2015, kim_2016, diaz_garcia_2016, algorry_2017}. This implies that, in the context of the bar continuum, bars will move over this continuum over time.

Other studies have suggested that there are differences in the surface brightness profile of weak and strong bars. Weaker bars tend to have exponential profiles, whereas stronger bars have flatter profiles \citep{elmegreen_1985, elmegreen_1996, kim_2015, kruk_2018}. Furthermore, \citet{kim_2015} found a correlation with bar length as well: longer bars, on average, have flatter profiles. \citet{kim_2015} speculate that bars initially have an exponential profile, which evolves over time (by trapping more stars into bar orbits) to a flatter profile. This allows for surface brightness profiles in between exponential and flat, which fits nicely with our hypothesis of the continuous nature of bar types and suggests that the surface brightness profile of the bar scales according to our proposed bar continuum.

\section{Conclusion}
\label{section:conclusion}
We have used the newest version of Galaxy Zoo, Galaxy Zoo DECaLS (GZD), to obtain a statistically significant and robust sample of \highlight{1,867} face-on disk galaxies with reliable bar classifications. The images used in GZD were obtained from the Dark Energy Camera Legacy Survey (DECaLS, \citealt{dey_2019}). Using the classification scheme shown in Table \ref{tab:bar_conditions}, we found that \highlight{28.1\%} of all our galaxies are classified as having a weak bar, while \highlight{15.5\%} are classified as having a strong bar. This results in a total barred fraction of \highlight{43.6}\%. We combined this data with SFR and stellar mass measurements from MPA-JHU \citep{kauffmann_2003,brinchmann_2004,tremonti_2004} and gas mass measurements from ALFALFA \citep{giovanelli_2005,haynes_2011,haynes_2018} to conduct a study looking at the differences between strong and weak bars for the first time.  

Our sample is consistent with the catalogue of visual morphological classifications of \citet{nair_2010} as well as with GZ2. We do measure a higher total barred fraction with GZD compared to GZ2, due to an improved decision tree and deeper imaging. Nevertheless, GZ2 did detect most GZD strong bars. Most of the newly detected bars in GZD (compared with GZ2) are weak bars. 

In addition, we saw that strong bars are longer than weak bars in terms of both the absolute and relative bar length. The average strong bars covers \highlight{40-45}\% of their host galaxy, whereas the average weak bar covers only \highlight{20-30}\%, although there is still significant overlap. Bars in quiescent galaxies are longer (in absolute length) than bars in SF galaxies for both weak and strong bars. 

However, most importantly, we have found that:
\begin{enumerate} 

    \item \textit{\textbf{The strongest bars facilitate quenching in SF galaxies.}}
    \begin{itemize}
        \item The strong bar fraction is higher in the red sequence and in quiescent galaxies (\highlight{24.9\% and 22.0\%}, respectively) than in the blue cloud and in SF galaxies (\highlight{9.6\% and 13.0\%, respectively}), whereas the weak bar fraction is roughly similar in both the red sequence and quiescent galaxies (\highlight{24.9\% and 25.0\%}, respectively) and the blue cloud and in SF galaxies (\highlight{29.9\% and 29.3\%, respectively})
        \item SF galaxies with a stronger bar have, on average, significantly higher fibre SFRs, lower gas masses and shorter depletion timescales than unbarred SF galaxies for a given stellar mass. This highlights that SF galaxies with stronger bars are usually rapidly evolving galaxies and that the strong bar facilitates the quenching process in these galaxies. These differences were not observed between SF galaxies with a weaker bar and SF unbarred galaxies.
        \item Quiescent galaxies with weaker and stronger bars do not significantly differ from quiescent galaxies with no bar.
    \end{itemize}

    \item \textit{\textbf{Weak and strong bars are part of a continuum of bar types.}}
    \begin{itemize}
        \item We have found that weak and strong bars are not fundamentally distinct physical phenomena, as we have shown that all differences between weak and strong bars disappear when controlling for bar length. Thus, there exists a continuum of bar types, for which weak and strong are the labels given to the extremes of this continuum.
        \item Nevertheless, it is still worthwhile to try to classify stronger and weaker bars into separate categories. This is because the effects that the bar exerts on its host depends on its position on the continuum. However, one must be cautious when doing so and be aware that results obtained by splitting bars into groups will only probe the extremes of the continuum.
    \end{itemize}

\end{enumerate}

\section*{Acknowledgements}

The data in this paper are the result of the efforts of the Galaxy Zoo volunteers, without whom none of this work would be possible. Their efforts are individually acknowledged at \url{http://authors.galaxyzoo.org}.

TG gratefully acknowledges funding from the University of Oxford Department of Physics and the Saven Scholarship.

RJS gratefully acknowledges funding from Christ Church, University of Oxford.

Funding for the Sloan Digital Sky 
Survey IV has been provided by the 
Alfred P. Sloan Foundation, the U.S. 
Department of Energy Office of 
Science, and the Participating 
Institutions. 

SDSS-IV acknowledges support and 
resources from the Center for High 
Performance Computing  at the 
University of Utah. The SDSS 
website is www.sdss.org.

SDSS-IV is managed by the 
Astrophysical Research Consortium 
for the Participating Institutions 
of the SDSS Collaboration including 
the Brazilian Participation Group, 
the Carnegie Institution for Science, 
Carnegie Mellon University, Center for 
Astrophysics | Harvard \& 
Smithsonian, the Chilean Participation 
Group, the French Participation Group, 
Instituto de Astrof\'isica de 
Canarias, The Johns Hopkins 
University, Kavli Institute for the 
Physics and Mathematics of the 
Universe (IPMU) / University of 
Tokyo, the Korean Participation Group, 
Lawrence Berkeley National Laboratory, 
Leibniz Institut f\"ur Astrophysik 
Potsdam (AIP),  Max-Planck-Institut 
f\"ur Astronomie (MPIA Heidelberg), 
Max-Planck-Institut f\"ur 
Astrophysik (MPA Garching), 
Max-Planck-Institut f\"ur 
Extraterrestrische Physik (MPE), 
National Astronomical Observatories of 
China, New Mexico State University, 
New York University, University of 
Notre Dame, Observat\'ario 
Nacional / MCTI, The Ohio State 
University, Pennsylvania State 
University, Shanghai 
Astronomical Observatory, United 
Kingdom Participation Group, 
Universidad Nacional Aut\'onoma 
de M\'exico, University of Arizona, 
University of Colorado Boulder, 
University of Oxford, University of 
Portsmouth, University of Utah, 
University of Virginia, University 
of Washington, University of 
Wisconsin, Vanderbilt University, 
and Yale University.

The Legacy Surveys consist of three individual and complementary projects: the Dark Energy Camera Legacy Survey (DECaLS; Proposal ID \#2014B-0404; PIs: David Schlegel and Arjun Dey), the Beijing-Arizona Sky Survey (BASS; NOAO Prop. ID \#2015A-0801; PIs: Zhou Xu and Xiaohui Fan), and the Mayall z-band Legacy Survey (MzLS; Prop. ID \#2016A-0453; PI: Arjun Dey). DECaLS, BASS and MzLS together include data obtained, respectively, at the Blanco telescope, Cerro Tololo Inter-American Observatory, NSF’s NOIRLab; the Bok telescope, Steward Observatory, University of Arizona; and the Mayall telescope, Kitt Peak National Observatory, NOIRLab. The Legacy Surveys project is honored to be permitted to conduct astronomical research on Iolkam Du’ag (Kitt Peak), a mountain with particular significance to the Tohono O’odham Nation.

The authors would like to thank the ALFALFA team for observing and processing of the ALFALFA data set.

This research made use of Astropy,\footnote{\url{http://www.astropy.org}} a community-developed core Python package for Astronomy \citep{astropy_2013, astropy_2018}, as well as other Python packages, such as Matplotlib \citep{hunter_2007}, NumPy \citep{harris_2020}, linmix \footnote{\url{http://linmix.readthedocs.org/}}, lifelines \citep{davidson_pilon_2021}, pandas \citep{mckinney_2010, pandas_2020} and SciPy \citep{virtanen_2020}.

This research made use of TOPCAT \citep{taylor_2005}.

\section*{Data availability}
The GZD data used in this article is made publicly available via Zenodo at https://dx.doi.org/10.5281/zenodo.4196267. The DECaLS DR8 data can be found at https://www.legacysurvey.org/dr8/. The latest SDSS data can be found at https://www.sdss.org/dr16/. The MPA-JHU VAC can be found at https://wwwmpa.mpa-garching.mpg.de/SDSS/DR7/. The ALFALFA data used in this article can be found at http://egg.astro.cornell.edu/.

\bibliographystyle{mnras}
\bibliography{bibtex.bib}
\appendix

\section{Comparison to Nair and Abraham}
\label{app:na}
\begin{figure}
	\includegraphics[width=\columnwidth]{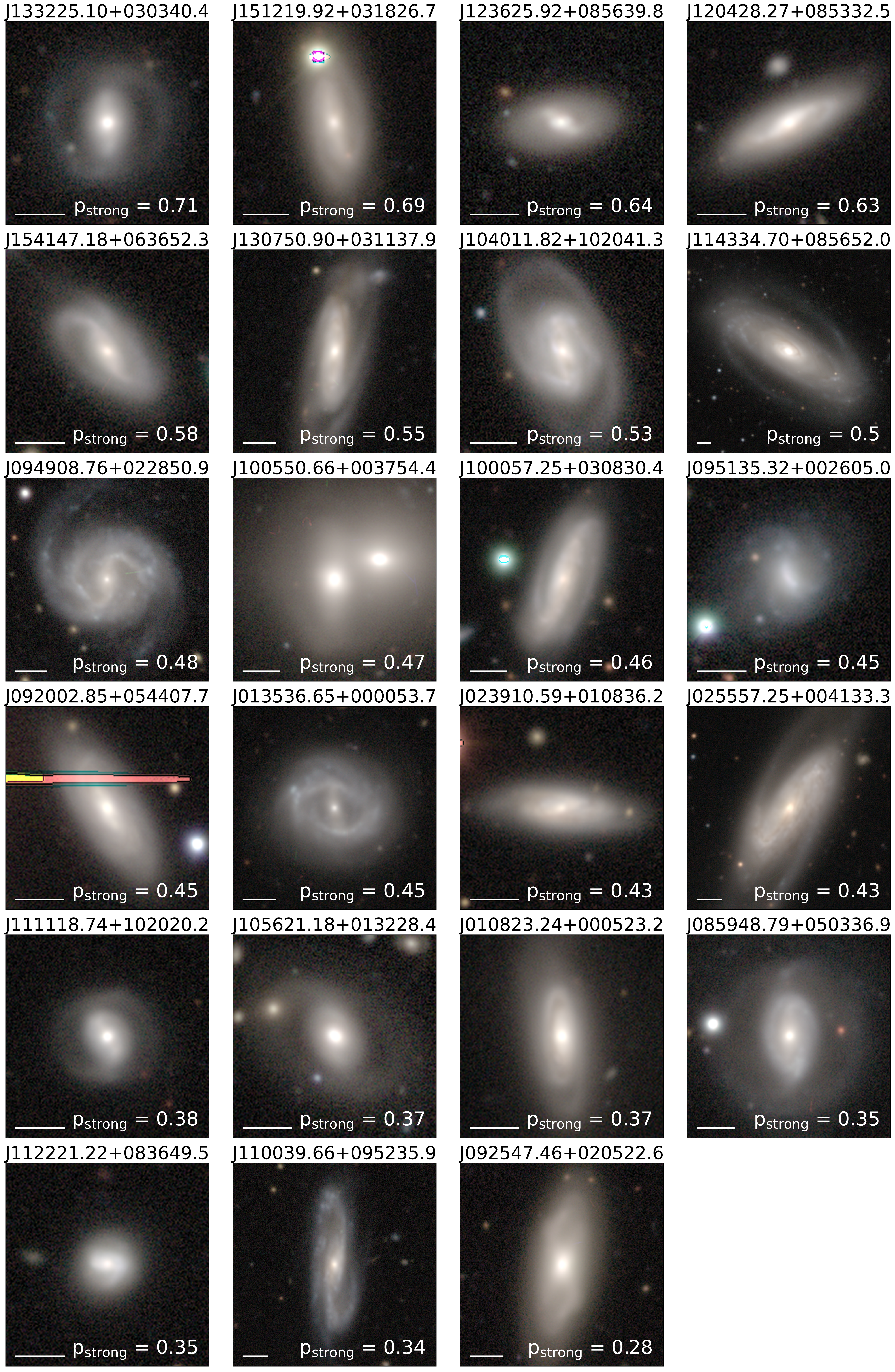}
    \caption{DECaLS postage stamps of the \highlight{23} galaxies classified by \citet{nair_2010} as unbarred, while GZD classified them as having strong bars. The white horizontal line in the bottom left corner of each subplot is 10 arcsecs long. The strong bar vote fraction ($p_{\textrm{strong bar}}$) of every galaxy is shown in the bottom right corner of each subplot. The galaxies are ordered from a high $p_{\textrm{strong bar}}$ (top left) to a low $p_{\textrm{strong bar}}$ (bottom right). The authors of this paper agree that the majority of these galaxies have a strong bar. The horizontal line over galaxy J092002.85+054407.7 is part of the mask of a very bright star just outside the frame.} 
    \label{fig:bars_GZDstrongNAno}
\end{figure}

We compared GZD to the catalogue of visual morphological classifications of \citet{nair_2010} in section \ref{subsubsection:na}. We found that there are \highlight{23} galaxies that \citet{nair_2010} has classified as unbarred, whereas the same galaxies were classified by GZD as have a strong bar. Those galaxies are visualised in Figure \ref{fig:bars_GZDstrongNAno}.

The authors of this paper agree that the vast majority of these galaxies have a strong bar. The reason for this disagreement between \citet{nair_2010} and GZD is most likely rooted in the different imaging used. GZD had access to DECaLS, while \citet{nair_2010} used SDSS images. 

\section{Effect of fibre size}
\label{app:fibre}

\begin{figure}
	\includegraphics[width=\columnwidth]{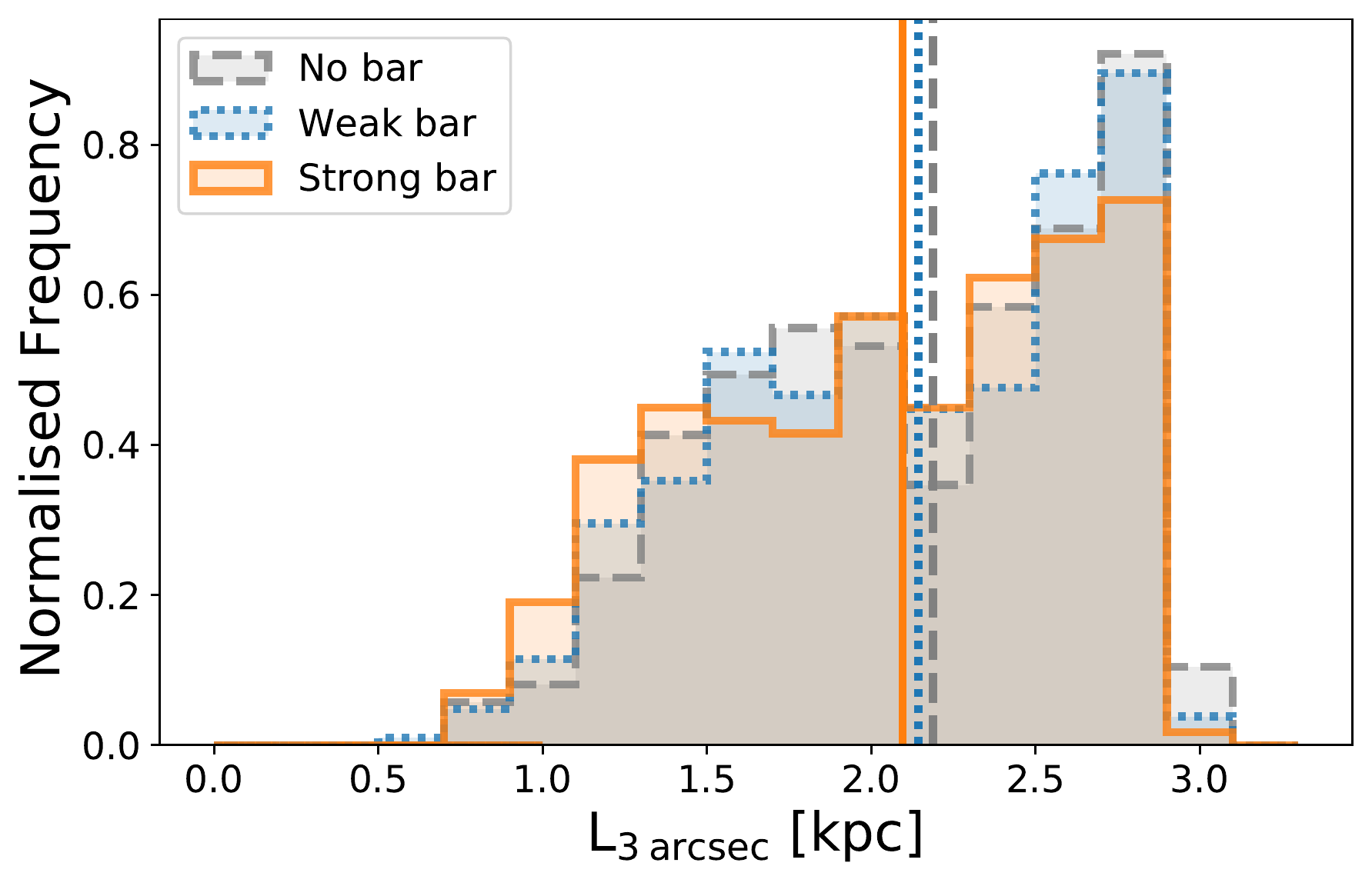}
    \caption{Histogram of the distribution of the physical distance corresponding to the 3 arcsec fibre (L$_{\textrm{3 arcsec}}$) of the three bar types. The median value for every bar type is indicated with a dashed line. The three distributions do not significantly differ from each other (the results of a KS tests for the 3 comparisons are always \highlight{< $1.5\sigma$).}} 
    \label{fig:d_3arcsec_hist}
\end{figure}

Throughout the paper, we have compared the fibre SFRs of galaxies with strong and weak bars. The fibre SFR probes the central 3 arcsec region of every galaxy. However, this 3 arcsec region will correspond to different physical distances depending on the redshift of the galaxy. We do not expect a change in the proportion of bar types over the short redshift range we are examining, but it is nevertheless worthwhile to investigate whether this will introduce a bias into our results. The distribution of the physical distance corresponding to the 3 arcsec fibre (L$_{\textrm{3 arcsec}}$) for every bar type is visualised in Figure \ref{fig:d_3arcsec_hist}. The three distributions do not differ significantly from each other (KS test: all three \highlight{< $1.5\sigma$)}.

\begin{figure*}
	\includegraphics[width=0.8\textwidth]{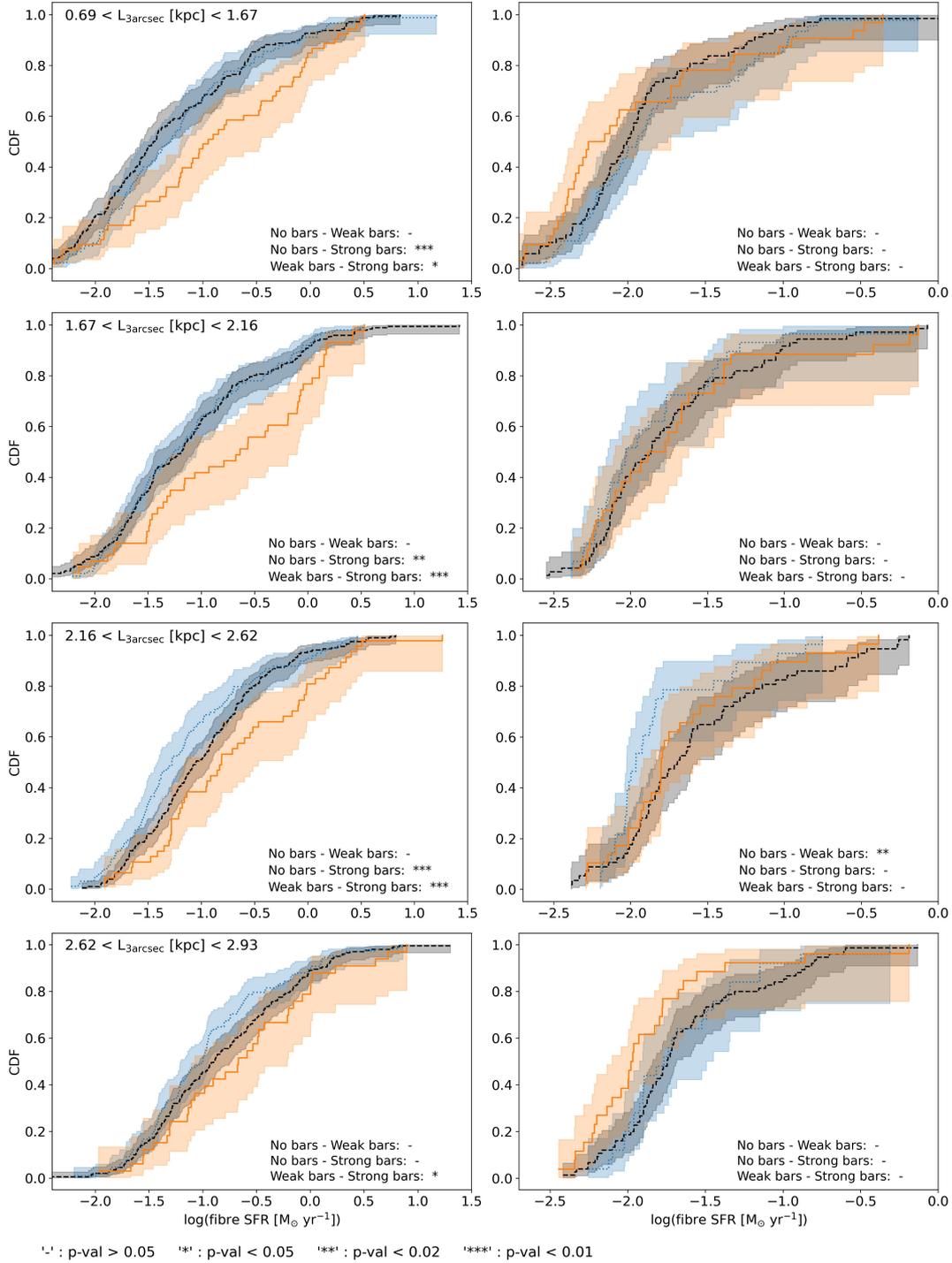}
    \caption{The results from a survival analysis on galaxies with no bar (black dashed) a weak bar (blue dotted) and a strong bar (orange full) in the SF group (left) and quiescent group (right). The galaxies are divided in four bins of L$_{\textrm{3 arcsec}}$ (i.e., physical distance corresponding to the 3 arcsec fibre in kpc). The top row displays the results for the lowest L$_{\textrm{3 arcsec}}$ bin and the bottom row the highest L$_{\textrm{3 arcsec}}$ bin. The results are shown in the form of cumulative density functions (CDFs). The results of the Cox's proportional hazards model are displayed in the bottom-right corner of every panel. If the p-value is below 0.01 (i.e., they are distinct), it is marked with `***'. If the p-value was between 0.01 and 0.02, it is denoted with `**'. If it was between 0.02 and 0.05, `*' was used. If it is above 0.05, it was denoted with `-'.}
    \label{fig:surv_fibre_d_3arcsec}
\end{figure*}

As a further test, we divided our sample into four L$_{\textrm{3 arcsec}}$ bins and recreated the survival analysis plots for fibre SFR (similarily to Figure \ref{fig:surv_comb}) for those different bins. The results are given in Figure \ref{fig:surv_fibre_d_3arcsec}. 

Again, the quiescent group (right column) does not exhibit much difference between the different bar types, while the SF group (left column) does. Most of the differences are observed between SF galaxies with strong bars and the other bar types. Interestingly, we can see that at the highest L$_{\textrm{3 arcsec}}$ bin (so at the highest redshifts), the differences are being `washed out' by the surrounding areas that are now within the fibre. Nevertheless, our overall conclusions remain the same.

\section{Co-evolution of bar and galaxy}
\label{app:coevolution}
In Section \ref{subsection:barlengths}, we suggested that the bar could co-evolve with the galaxy, based on the bar length data shown in Figure \ref{fig:barlen}. We saw that a strong bar in a SF galaxy tends to be shorter than a strong bar in a quiescent galaxy, and suggested that the bar co-evolves with its host and is still growing when in a SF galaxy and is fully grown once the host is quenched. Similar results were found for weak bars.

However, it is also possible that this result is obtained as direct consequence of the fact that volunteers classify based on the relative length of the bar, as instructed, and that quiescent galaxies tend to be bigger than SF galaxies. 

\begin{figure}
	\includegraphics[width=\columnwidth]{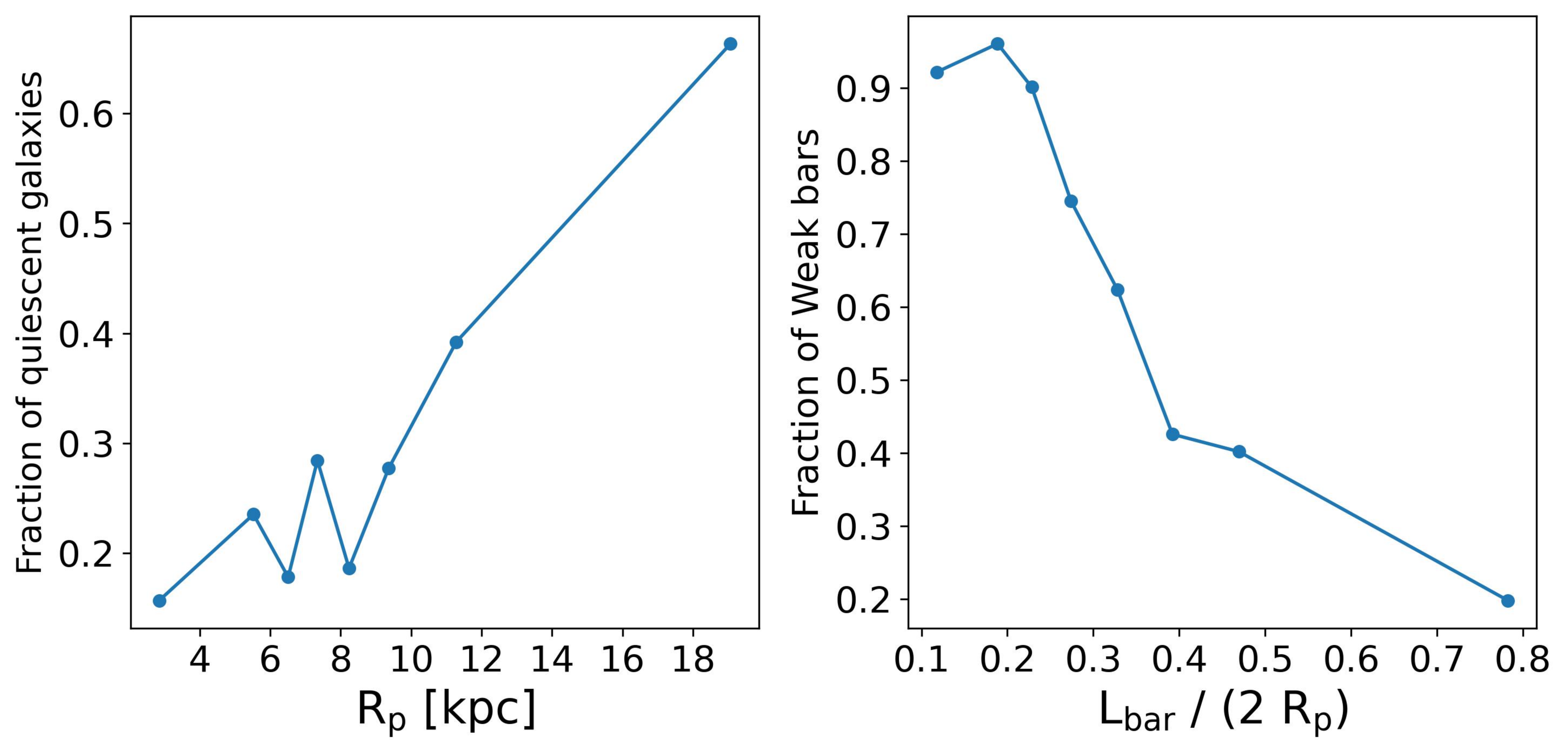}
    \caption{\textbf{Left}: The fraction of quiescent galaxies over Petrosian radius (R$_{\rm p}$) in eight equal-size bins. These fractions are used for reassigning the quiescent/SF labels to the galaxies. \textbf{Right}: The fraction of weak bars over the relative size of the galaxy. These fractions are used for reassigning the weak/strong bar labels.}
    \label{fig:co_evolution_bins}
\end{figure}

This idea can be tested by temporarily stripping the weak/strong and quiescent/SF labels from each galaxy, and redistributing them according to the following rules: 1) quiescent galaxies are bigger and 2) strong bars are longer in terms of the relative size of the bar. To be more accurate, our sample was binned in eight equal-sized bins of Petrosian radius and for every bin we calculated what fraction of galaxies were quiescent (left plot of Figure \ref{fig:co_evolution_bins}). Then, when reassigning labels, the chance of labelling a particular galaxy to be `quiescent' was equal to that fraction. This was done as well for the relative bar length and the strong/weak bar labels (right plot of \ref{fig:co_evolution_bins}).

\begin{figure}
	\includegraphics[width=\columnwidth]{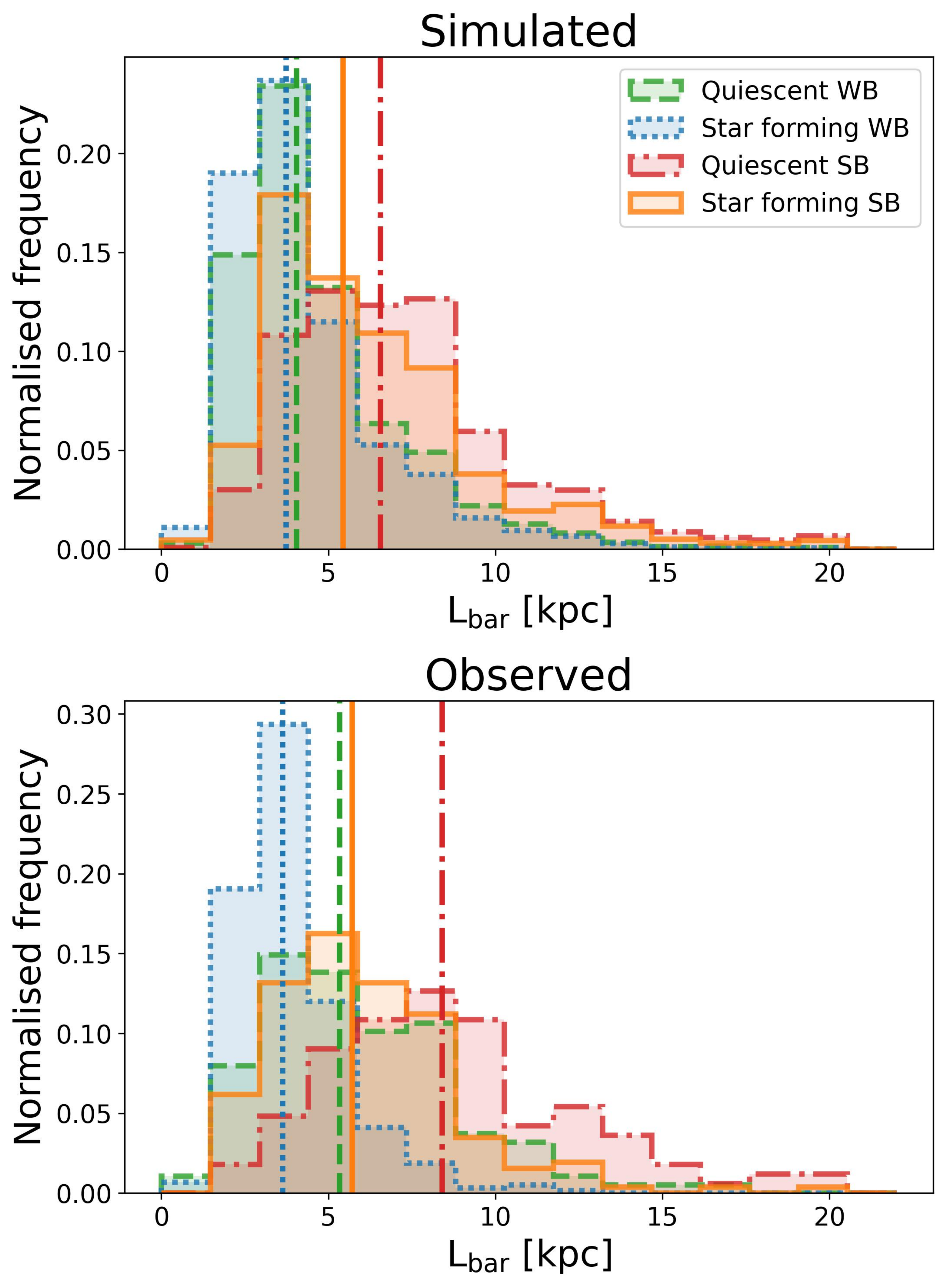}
    \caption{\textbf{Top}: The distribution of absolute bar lengths for the 1,000 different simulations for the four different subsamples: quiescent galaxies with a weak bar, quiescent galaxies with a strong bar, SF galaxies with a weak bar and SF galaxies with a strong bar. \textbf{Bottom}: The observed absolute bar lengths, identical to the left plot of Figure \ref{fig:barlen}. As in the observed data, we see that a strong bar in a SF galaxy tends to be shorter than a strong bar in a quiescent galaxy in the simulation (and similarily for weak bars). However, the difference is bigger in the observed data than in the simulated data.} 
    \label{fig:co_evolution}
\end{figure}

The result of doing this a 1,000 times per galaxy and plotting the bar length distribution for our four subsamples (quiescent galaxies with a weak bar, quiescent galaxies with a strong bar, SF galaxies with a weak bar and SF galaxies with a strong bar) is given in the top plot of Figure \ref{fig:co_evolution}. As expected, the same trends that we saw in the observed data are found: a strong bar in a SF galaxy tends to be shorter than a strong bar in a quiescent galaxy (and similarily for weak bars). Although, the difference is not as large in the simulated results as in the observed results, shown in the bottom plot of Figure \ref{fig:co_evolution}. Additionally, none of the observed distributions are likely drawn from their corresponding simulated distribution (p-value < \highlight{0.005} from a KS-test), except for the SF galaxies with a strong bar. Thus, there is still an unexplained effect, which is consistent with (but does not prove) the co-evolution theory described in Section \ref{subsection:barlengths}. 

\bsp
\label{lastpage}
\end{document}